\documentclass[preprint,showpacs,preprintnumbers,amssymb,nofootinbib,superscriptaddress,aps]{revtex4-2}
\pdfoutput=1
\usepackage{amsmath,amssymb,bm}
\usepackage{bbold}
\usepackage{lipsum}
\usepackage{hyperref}
\usepackage{graphics,graphicx}
\usepackage{subcaption}
\usepackage{csquotes}
\usepackage{mathtools}
\usepackage{slashed}
\usepackage{color,colordvi}
\usepackage{soul}
\usepackage{xcolor}
\usepackage{array}
\usepackage{multirow}
\usepackage{silence}
\usepackage{booktabs}
\usepackage{multirow}
\usepackage{tabularx,ragged2e}
\usepackage{comment}
\usepackage{slashed}
\usepackage{float}
\usepackage{epigraph}
\usepackage[normalem]{ulem}

\renewcommand{\{}{\left\lbrace}
\renewcommand{\}}{\right\rbrace}

\newcommand{\GeV}{\,\mathrm{GeV}}

\newcommand{\MeV}{\,\mathrm{MeV}}

\begin{document}
   
\title{Chiral Perturbation Theory\texorpdfstring{\\Reflections on Effective Theories of the Standard Model}{}}
\author{B. Ananthanarayan}
\email{anant@iisc.ac.in}
\affiliation{%
 Centre For High Energy Physics, Indian Institute of Science\\
        Bangalore 560 012, India. }
\author{M. S. A. Alam Khan}%
 \email{mohdakbar@iisc.ac.in}
\affiliation{%
 Centre For High Energy Physics, Indian Institute of Science\\
        Bangalore 560 012, India. }%

\author{Daniel Wyler}
\email{wyler@physik.uzh.ch}
\affiliation{
 Institute for Theoretical Physics
        University of Z\"urich\\
        Winterthurerstr. 190, CH 8057 Z\"urich, Switzerland.}%

 \begin{abstract}
The pseudoscalar particles pions, kaons and the $\eta$-particle are considerably lighter than the other hadrons such as protons or neutrons. Their lightness was understood as a consequence of approximate chiral symmetry breaking. This led to current algebra, a way to express the relations imposed by the symmetry breaking. It was realized by Weinberg that because of their low mass, it is possible to formulate a purely pionic (effective) field theory at experimental energies,  which carries all information on the (non-perturbative) dynamics, symmetries, and their spontaneous breaking of quantum chromodynamics (QCD) and allows for systematic calculations of observables. In this review, we trace these developments and present recent activities in this field. We make the connection to other effective theories, more generally introduced by Wilson, as approximate field theories at low energies. Indeed, principles and paradigms introduced first for pions have become ubiquitous in particle physics and the standard model. Lastly, we turn to the latest development where the present (fundamental) standard model itself is considered as an effective field theory of a - yet to be formulated - even more fundamental theory. We also discuss important techniques that were developed in order to turn chiral perturbation theory into a predictive framework and briefly review some connections between lattice QCD and chiral perturbation theory (ChPT).
\end{abstract}
\maketitle


\tableofcontents
\newpage
\begin{widetext}
  \setlength{\epigraphwidth}{0.9\textwidth}
     \epigraph{ \it{``And so the question naturally arose, is there a way of avoiding the machinery of current algebra by just writing down a field theory that would automatically produce the same results with much greater ease and perhaps physical clarity? Because after all in using current algebra one had to always wave one’s hands and make assumptions about the smoothness of matrix elements, whereas if you could get these results from Feynman diagrams, you could see what the singularity structure of the matrix elements was and make only those smoothness assumptions that were consistent with that."}}{Steven Weinberg, 2020\cite{Weinberg:2021exr}}
     \end{widetext}
    \section*{Preamble}
       We dedicate this article to Steven Weinberg, a great physicist who influenced in many ways the field of elementary particle physics for decades and is one of the authors of the Standard Model. Apart from his great achievements in research, his textbooks which became classics also testify to his outstanding teaching capacity. He was awarded the 1979 Nobel Prize in physics along with Sheldon Glashow and Abdus Salam for the electro-weak model.  In addition, Weinberg pioneered the study of quantum field theories, behavior of Green functions at asymptotic energies, symmetries in field theories, Goldstone mechanism, current algebra, pion physics and effective field theories. Weinberg has explained his philosophy in terms of phenomenological Lagrangians~\cite{Weinberg:1978kz}, which proved to be the cornerstone for successful developments in precision pion physics until today.  This concept of an effective Lagrangian, that is, the low energy (large distance) manifestation of a fundamental theory, was also developed by Ken Wilson~\cite{Wilson:1969zs} (Nobel prize 1982) quite generally. In this picture, short-distance degrees of freedom are systematically `integrated out' and appear only as coefficients of a theory with long-distance degrees of freedom. Weinberg's and Wilson's concepts are at the base of today's understanding of physics systems with very different energy scales, such as in particle physics where the range goes over more than 20 orders of magnitudes~\cite{Wells:2012rla}.\par
    In this review, we aim to showcase the developments and richness of effective theories, report on current progress, and encourage further work by pointing out where it is needed. We mention important early work quite comprehensively but are more anecdotal in referencing newer work; the interested reader should be able to navigate it from the references given.\par 
We have striven to bring under one umbrella several topics that have been separately reviewed for various sub-communities. Our hope is that the present review will find a readership that will encompass all of the working particle physics community, experimentalists and theorists likewise, who wish to get a flavor of what has been going on under the rubric of chiral perturbation theory and a glimpse of other effective field theories. The subject of chiral perturbation has grown immensely based on rather technical and detailed computations, in this review, we wish not to burden the text with too many equations but try to explain the physical concepts and point the interested reader to more comprehensive reviews. In particular, we illustrate how the general principles of analyticity, unitarity, and crossing (through dispersion relations and the analysis of experimental data based on them), can be combined with the scattering amplitudes arising in chiral perturbation theory. This marriage has in fact led to sufficient accuracy, thereby providing for testing the standard model at requisite levels of precision. In addition, we have also given references to the various packages used in the literature which readers can find interesting.
    \section*{A personal note}
     \noindent
     Both DW and BA have been working for over three decades on the subjects discussed here, in particular in pion physics and Chiral Perturbation Theory (ChPT). We are pleased to share the important lessons learned during the many years of development in the field. DW was happy to contribute and to assist BA who has been a longtime friend and a gate opener to India and its culture.   
     
    \section{Introduction}
    \label{sectionINT}
    Many physical systems look very different when probed at different length scales or with different energies. While ordinary matter appears to the eye in an incredibly rich diversity of forms and textures, at the atomic scale, made for instance visible by scanning microscope techniques, all one sees are atoms that are quite similar in different materials. Thermodynamics, the phenomenological description of many systems, can be viewed as an effective `leftover' of the microscopic theory of statistical mechanics. Such large differences appear in many physical systems.  In elementary particle physics, it is in the realm of strong interactions where this can be studied particularly well. At experimental energies beyond, say, several $\GeV$, the relevant picture is that of the simple $SU(3)$ gauge theory of QCD. But at lower energies, the complicated interactions of pions and nucleons dominate and there is no obvious trace of QCD. While we consider QCD the fundamental theory, the interactions of pions and nucleons are described by an effective (low energy) theory, called chiral perturbation theory, ChPT.  So, how does one connect these two seemingly different manifestations of the same interactions?\par
    The key is to find properties of QCD that remain manifest also in ChPT.  In this case, it turned out that the crucial property is chiral symmetry $SU(3)\times SU(3)$ with spontaneous breaking and a small explicit breaking term. While this symmetry is easily gleaned from the fundamental Lagrangian of QCD, it is far less obvious at the effective level. It took many years, from the late 1950s on, to consolidate the effects of that symmetry, and many physicists are associated with this process. Since these developments are a fascinating part of the history of particle physics, we will summarize some of these ideas below.

    All the results obtained were cast into a bona fide field theory using external field techniques by Gasser and Leutwyler in the 1980s. Apart from a rigorous formulation of the effective Lagrangian, they gave a complete one-loop treatment of the effective Lagrangian, also including the other pseudoscalar mesons of the eightfold way, that is the kaons and the $\eta$. Since then, many developments have taken place. Systematic two-loop calculations increased to quality of the predictions substantially. The role of nucleons and vector mesons was investigated; general techniques such as dispersion relations, functional analysis methods, and rescattering theory advanced the precision of the calculations. Other methods, such as lattice gauge theory, furnished important input. All of these represent diverse and rich activities in physics where each would require a review article in its own right. Furthermore, the success of chiral perturbation theory encouraged the development of many other effective theories in particle physics. By identifying high energy and low energy degrees of freedom and using Wilson's procedure to integrate out the high energy modes, it is possible to arrive at an effective theory for the low energy modes in many cases. Some of the recent theories are the heavy quark effective theory (HQET) and the soft-collinear effective theory (SCET). For instance, HQET is based on the observation that in heavy quark physics (mostly $b$ quarks but also charmed quarks) mesons, in the limit of large (bigger than about $1\GeV$) heavy quark mass $m_q$, the relevant physics can be reliably described in a power series in $(m_q)^{-1}$ where the first term is independent of $(m_q)$ \footnote{Recall that the reduced mass of a system of a heavy and a light particle is largely independent of the heavy mass}. \par
    
    While these theories are designed to understand the non-perturbative dynamics of QCD better, more recently, the idea that even the `fundamental' standard model is but the `low' energy effective manifestation of a more basic theory that would reveal itself at very high energies. This idea is known as the standard model effective theory (SMEFT). A further step is to view the theory of gravity, Einstein's general relativity, as an effective theory. This is particularly interesting for attempts to turn gravity into a quantum theory.\par 
    
    The contents of this review are as follows: 
    
    The next section \ref{sectionCL} gives an overview of chiral perturbation theory. We recall the basic theory of QCD and describe the construction of the effective Lagrangian, following Gasser Leutwyler~\cite{Gasser:1984gg}. Several general observations, sometimes personal, are mixed into the text.
    
    In section \ref{ECT}, we show how to include particles beyond the eight light pseudoscalar mesons, in particular the $\eta'$, the vector mesons (such as the $\rho$), and nucleons (baryons).
    
    In section \ref{TBS}, we consider processes where ChPT does not work well and must be improved by auxiliary methods. In particular, we look here at strong two- and three-body rescattering where a substantial body of work exists.
    
    In section \ref{GRG}, we look at generalizations of the renormalization procedure for non-renormalizable theories. The methods that have been developed might be of general
    interest in going beyond `renormalizable' theories.
    
    In section \ref{Weak} we review the weak interactions of the pseudoscalar mesons, such as the decays of kaons which play an important role in the understanding of fundamental effects such as CP-violation.
    
     In section \ref{APP}, we show some applications which are of special importance.
     
      In section \ref{OSI}, we discuss effective methods to deal with the strong interactions in the higher energy regimes where ChPT does not apply (or only in certain parts of phase space).
      
      Then, in section \ref{SMEFT}, we show the newest development in understanding the standard model and gravity as effective theories of an even more fundamental theory.

      Finally, in section \ref{MISC}, we collect some noteworthy recent developments that are of relevance for ChPT.

   \section{The Chiral Lagrangian}
   \label{sectionCL}
  After the initial work of Dashen, Weinstein, and Pagels~\cite{Dashen:1969eg,Dashen:1970et, Dashen:1969ez, Pagels:1974se}, a breakthrough came from the
observations of Weinberg~\cite{Weinberg:1978kz} who argued based on the principles of quantum field theory and cluster decomposition and pion-pole dominance, that the lowest order effective Lagrangian
could be used to compute loops whose divergences could be absorbed into the low-energy constants of higher order terms in the Lagrangian. This was put on a firm footing by studying
the gauge invariance of the generating functional of the Green functions of the theory by
Gasser and Leutwyler~\cite{Gasser:1983yg, Gasser:1984gg}.  A scholarly exposition is given in the Scholarpedia article of Leutwyler~\cite{Leutwyler:scpedia}.
 In this chapter, we review their construction of the chiral Lagrangian.\hfill\par
    The foundation for this is QCD, the theory of the strong interactions which determine the behavior of the observed particles in a variety of experiments where they are influenced by `external fields' or `sources'. These are classical objects and do not appear in loops. Important examples of such external fields can be the masses of the quarks\footnote{In the standard model this is proportional to the vacuum expectation value of the Higgs field.}, the weak interactions, or also experimentally realized fields like a strong electromagnetic field. Since we know that the symmetry properties are crucial, we are interested in external fields that have well-defined transformation properties under the chiral symmetry which determines the low energy spectrum. The objects one wants to calculate are the Green functions associated with the external fields, from which the physical matrix amplitudes are derived in a standard manner.

Thus, the fundamental Lagrangian involving the three light quark fields ($q$) has the form:
   \begin{equation}
        \mathcal{L}=\mathcal{L}_{Q C D}^{0}+\bar{q} \gamma^{\mu}\left(v_{\mu}+\gamma_{5} a_{\mu}\right) q-\bar{q}\left(s-i \gamma_{5} p\right) q - \frac{\theta}{32\pi^2}\text{Tr} \left(G_{\mu\nu}\tilde{G}^{\mu\nu}\right)\,,
          \label{anomaly}
    \end{equation}
where \begin{align}
     \mathcal{L}_{QC D}^{0}=\overline{q}i \gamma^{\mu}\left( \partial_{\mu}-i G_{\mu} \right) q -\frac{1}{2g^2}\text{Tr}\left(G^{\mu\nu}G_{\mu\nu}\right),
     \label{eq:eq2}
\end{align}
and $G^{\mu\nu}$ is the gluon field strength tensor and $\tilde{G}^{\mu\nu}=\frac{1}{2}\epsilon^{\mu\nu\alpha\beta}G_{\alpha\beta}$ its dual. 
The letters $v_\mu$, $a_\mu$, $s$ and $p$ denote the external fields transforming as vectors, axial vectors, scalars and pseudoscalars, respectively; the field $\theta$ transforms
in a particular non-linear way. All these quantities are $x$-dependent, that is $v_\mu = v_\mu(x)$, etc. The physical Greens functions from the Lagrangian in eq.~\eqref{anomaly} are obtained by expanding the generating function around $v
_{\mu}=a_{\mu}=p=0$, $s=\mathcal{M}$, $ \theta=\theta_0$ where $\mathcal{M}$ is the quark mass matrix and $\theta_0$ is the vacuum angle. We note that $s=\mathcal{M}$ can always be chosen to be diagonal, with real positive elements $\left(m_u, m_d, m_s\right)$ and an adjusted vacuum angle. For more details, see ref.~\cite{Gasser:1984gg}. The last term in eq.~\eqref{anomaly} is odd under the CP transformation and contributes  
to CP violation effects (for instance the electric dipole moment of the neutron)  from the strong interaction. These effects are found to be tiny, which requires the vacuum angle 
to be unnaturally small $\theta_0\lesssim10^{-10}$~\cite{Abel:2020pzs,Addazi:2022whi}. This is the, still unresolved, `strong CP'
problem. It has sparked many ideas, including the postulation of the axion, an interesting, but still hypothetical particle. Various theoretical models of axions and the experimental bounds on the couplings with other particles have been explored in the literature and these developments can be found in ref.~\cite{DiLuzio:2020wdo}, for a
very recent result, see ref.~\cite{Schulthess:2022pbp}.
\par

    Our interest is in the amplitudes at experimental particle energies below $1 \GeV$ or so. However, the calculations using the formulas above would be forbiddingly difficult because at low energies, there are no free quarks (or inclusive states, such as exist at high energies), but pions or kaons (the pseudoscalar mesons), or other
    hadrons that are complicated bound states of quarks and gluons.  Thus, we must express the physical contents of the QCD Lagrangian in terms of these fields. This we shall call the effective (low-energy) chiral Lagrangian. The notion is quite general: An effective Lagrangian expresses the physics in terms of the physical particles (or fields) at the energies relevant for the experiment considered.
        The form of the chiral Lagrangian is dictated by the choice of the physical (dynamical) fields and the symmetry properties of the external fields. Because we are interested in low energies, one considers an expansion in energy (momentum) of the particles which, because of chiral symmetry, starts at order($p^2$) where $p$ is a typical momentum of the particles. The leading term in an energy (momentum) expansion is completely fixed due to the work of Callan, Coleman, Wess and Zumino~\cite{Coleman:1969sm, Callan:1969sn} have allowed us to extract several general features of the interactions of Goldstone bosons, quite independent of the knowledge of the dynamics of the strong interactions, based exclusively on the (global) symmetries of the underlying Lagrangian. The fact that they are (approximate) Goldstone bosons already fixes their mutual interactions to be of the derivative type. Furthermore, the parametrization of the degrees of freedom encoded by the physical fields requires them to be coordinates of the coset space given by $G/H$, where $G$ is the global symmetry of the Lagrangian and $H$ is the symmetry of the ground state.  The broken generators of $G$ not lying in $H$ are precisely these Goldstone boson degrees of freedom. Taking into account the symmetry properties of the Goldstone bosons, a convenient parametrization where the chiral transformations are linear and ensures the derivative nature of the interactions is~\cite{Boulware:1981ns}:
      \begin{align}
        U\equiv e^{i \sqrt{2} \Phi/F_{\pi}}
        \label{eq:chiral_field}
    \end{align}
    where
    \begin{equation}
        \Phi=\left(\begin{array}{ccc}
            \frac{\pi^{0}}{\sqrt{2}}+\frac{\eta_{8}}{\sqrt{6}} & \pi^{+} & K^{+} \\
            \pi^{-} & -\frac{\pi^{0}}{\sqrt{2}}+\frac{\eta_{8}}{\sqrt{6}} & K^{0} \\
            K^{-} & K^{0} & -\frac{2 \eta_{8}}{\sqrt{6}}
        \end{array}\right)\,,
    \end{equation}
 which is unique up to the reparametrization of the Goldstone boson fields themselves. As argued by Boulware and Brown~\cite{Boulware:1981ns}, it is advantageous to group the pseudoscalar fields into a $3\times3$ unitary. 
The point is that the external fields should couple to such operators of the fields which transform linearly under chiral transformations in order that the interactions can be built by a conventional loop expansion.

    The results of current algebra are all captured by the effective Lagrangian:
    \begin{align}
        \mathcal{L}_{eff}=\mathcal{L}_{2}+\mathcal{L}_{4}+\mathcal{L}_{6}+\cdots
    \end{align}

  where $\mathcal{L}_{2}$, $\mathcal{L}_{4}$ and $\mathcal{L}_{6}$ are the terms of  $\mathcal{O}\left(p^2\right)$, $\mathcal{O}\left(p^4\right)$ and $\mathcal{O}\left(p^6\right)$, respectively.

    The leading-order term in the low energy expansion is generated by the non-linear sigma model coupled to the external fields(v, a, s, and p, in the notation of the ref.~\cite{Maiani:1995ve}),
    \begin{equation}
        \mathcal{L}_2=\frac{F_{\pi}^2}{4}\langle D_{\mu}UD^{\mu}U^{\dagger}+\chi U^{\dagger}+\chi^{\dagger}U\rangle\,,
    \end{equation}
    where
    \begin{equation}
        D_{\mu} U=\partial_{\mu}U-i (v_{\mu}+a_{\mu})U+i U (v_{\mu}-a_{\mu}),\quad \chi=2B\left(s+i p\right)\,.
    \end{equation}
    
The above reproduces the well-known Weinberg result for $\pi\pi$ scattering~\cite{Weinberg:1966kf}, which sets the scale of the chiral interaction in terms of the pion decay constant.    

Without dynamic external fields, $s$ is proportional to the masses
    of the quarks, which are fundamental quantities in the standard model.
    In fact, ChPT plays a key role in determining these quantities in terms of the pseudoscalar masses~\cite{Gell-Mann:1968hlm,Halprin:1976rs,Weinberg:1977hb,Gasser:1984gg}.
    There is also considerable effort to determine them from lattice QCD~\cite{Aoki:2021kgd} as well as using QCD sum rules~\cite{Shifman:1978bx,Shifman:1978by}. The sum 
 rule determinations are also a powerful tool to determine the QCD parameters, and their recent applications can be found in the book of Dominguez ~\cite{Dominguez:2018zzi}. A summary of the most recent determinations can be found in the PDG~\cite{Workman:2022ynf}.
 Apart from their importance as fundamental parameters, a value of $m_u = 0 $ would have
    solved the so-called strong CP problem (see~\cite{Hook:2018dlk} for details), but this does not seem to be the case.
    
    In particular, in order to go beyond the leading order, Weinberg
    in ref.~\cite{Weinberg:1978kz} argued that the content of a field theory is dictated by the
    symmetries and analyticity, perturbative unitarity, and cluster decomposition.  Using
    this, he was able to predict the structure of the $\pi\pi$ scattering amplitude and the corresponding logarithms that would have to generate the required imaginary parts of the amplitude from the original (real-valued) tree-level amplitude~\cite{Weinberg:1978kz}. While this remained a thumb rule, the systematic study required the introduction of external sources for the currents of the theory in
    the spirit of Julian Schwinger, who proposed to study field theory through sources. Rather than being mere mathematical curiosities, the presence of the external sources allowed a systematic computation of one-loop generating functional through the heat-kernel technique, an established method of obtaining in a compact manner the generating functional rather than an equivalent yet explicit computation of Feynman diagrams.  This functional requires regularization (dimensional) and renormalization
    of the infinities that are generated by the loops.
    
  Since the original Lagrangian in $d=4$ is not renormalizable, the procedure generates higher derivative terms not present in the original two-derivative Lagrangian.  Thus new low-energy constants are introduced into the theory with corresponding $\beta$-functions, which are fixed from the experiment.  Once fixed, at this order, any process of interest can be computed, which makes the theory predictive. This process can be continued indefinitely to any order in the loop expansion and/or the momentum expansion. It is the convention to consider each loop to yield a new power of $p^2$ and also to assign powers to the explicit mass.

   The chiral Lagrangian at $\mathcal{O}(p^4)$~\cite{Gasser:1984gg,Gasser:1983yg} consistent with the Lorentz invariance, C and P symmetry is given by:
    \begin{align}
         \mathcal{L}_4 = & L_1 \langle D_\mu U^\dagger D^\mu U\rangle^2 +L_2 \langle D_\mu U^\dagger D_\nu U\rangle \langle D^\mu U^\dagger D^\nu U\rangle\nonumber\\ &+ L_3 \langle D_\mu U^\dagger D^\mu U D_\nu U^\dagger D^\nu U\rangle +  L_4 \langle D_\mu U^\dagger D^\mu U\rangle \langle \chi^\dagger U +  \chi U^\dagger\rangle  \nonumber\\&+L_5 \langle D_\mu U^\dagger D^\mu U(\chi^\dagger U + U^\dagger \chi)\rangle+L_6 \langle \chi^\dagger U + \chi U^\dagger \rangle^2 +   L_7 \langle \chi^\dagger U - \chi U^\dagger \rangle^2  \nonumber\\& + L_8 \langle \chi^\dagger U \chi^\dagger U +\chi U^\dagger \chi U^\dagger\rangle -i L_9 \langle F_R^{\mu\nu} D_\mu U D_\nu U^\dagger + F_L^{\mu\nu} D_\mu U^\dagger D_\nu U \rangle \nonumber\\ &+ L_{10} \langle U^\dagger F_R^{\mu\nu} U F_{L\mu\nu}\rangle +  L_{11} \langle F_{R\mu\nu} F_R^{\mu\nu} + F_{L\mu\nu} F_L^{\mu\nu}\rangle + L_{12} \langle \chi^\dagger \chi \rangle\,,\label{eq:L4}
\end{align}
where
\begin{align}
  F_R^{\mu\nu} = & \partial^\mu r^\nu -\partial^\nu r^\mu -i [r^\mu,r^\nu]  \\
  F_L^{\mu\nu} = & \partial^\mu l^\nu -\partial^\nu l^\mu -i [l^\mu,l^\nu]\,.
    \end{align}
    At this order in the momentum expansion, there are twelve new couplings $L_i$ (also called low energy constants) that appear out of which only $L_3$ and $L_7$ are not divergent. \par
    
    Also, the next $\mathcal{O}\left(p^6\right)$ order Lagrangian has been worked out, see refs.~\cite{Fearing:1994ga,Bijnens:1999sh} and there are results of $\mathcal{O}\left(p^8\right)$~\cite{Bijnens:2017wba,Bijnens:2018lez}. Of course, the number of couplings increases, and thus the predictive power for smaller (higher-order) effects decreases. But by choosing suitable observables, one is still able to determine some of them.
    
    The low energy constants can be renormalized in a standard fashion and one writes:
\begin{align}
    L_i=L_i^r(\mu)+\Gamma_i \lambda(\mu)
    \label{eq:eq12}
\end{align}

where
\begin{align} 
\lambda(\mu)&=\frac{\mu^{d-4}}{4\pi^2}\left(\frac{1}{d-4}-\frac{1}{2}\left(1+\log(4\pi)-\gamma_E\right)\right)\,,
\end{align}
where the function $\lambda$ comes from performing a standard one loop calculation and the coefficients $\Gamma_i$ coefficients were calculated in~\cite{Gasser:1984gg,Gasser:1983yg}. The divergences in the bare coupling $L_i$ cancel with the one present in the $\lambda(\mu)$ resulting in renormalized couplings $L^r_i(\mu)$. Their scale dependence is given by:
    \begin{align}
    L^r_i(\mu_2)=L^r_i(\mu_1)+\frac{\Gamma_i}{16\pi^2}\log\left(\frac{\mu_1}{\mu_2}\right)\,.
     \label{eq:eq13}
     \end{align}

    The low energy constants reflect the properties of the strong interaction spectrum that has been integrated out and contribute to their numerical values. For instance, in certain cases, the exchange of a single vector meson accounts for the observed values; alternatively, there might be an axial-vector meson dominance~\cite{Ecker:1988te,Ecker:1989yg}. Other higher angular momentum states also make contributions but are mostly numerically less significant. Over the last years, several papers have explored ways to determine the low energy constants, see ref.~\cite{Bijnens:2014lea}. For a recent review, see ref.~\cite{Jiang:2022gjy}.

 To illustrate the renormalization procedure, we consider the  electromagnetic form factor of pion which  is defined as:
    \begin{equation}
        \langle\pi^+(p')\vert V_{\mu}^{em}(0)\vert \pi^+(p)\rangle=(p'+p)_{\mu}F^{\pi}_V(t),\quad t=(p'-p)^2\,.
    \end{equation}
    This quantity at tree-level and at one-loop is given by:
    \begin{align}
        F^{tree}_V(t)&=1\,,\\
        F^{\pi,1-loop}_V(t)&=2\phi(t,M_{\pi};d)+\phi(t,M_{K};d)\;,
    \end{align}
    where
    \begin{align}
        \phi(t,M;d)=-\frac{t M^{d-4}}{(4\pi)^{d/2}}\frac{\Gamma(2-d/2)}{2F_{\pi}^2}\int_0^1 dx \hspace{1mm}x(1-2x)\left(1-\frac{ t}{M^2}x(1-x)\right)^{\frac{d-4}{2}}
        \label{eq:phi_p2}
    \end{align}

The $\mathcal{O}(p^4)$ contribution to $F_V^{\pi}$ comes from $L_9$ and has the following form:
    \begin{align}
        F_V^{\pi,\mathcal{L}_4}=\frac{2L_9 t}{F_\pi^2}\,,
    \end{align}
    and the total contribution to $F_V^{\pi}$ is finite at this order if the $L_9$ is tuned as:
    \begin{align}
        L_9&=L_9^r(\mu)+\frac{\lambda(\mu)}{4}\,,\\
        \end{align}
    this results into finite scale dependent renormalized coupling $L_9^r(\mu)$. Now the divergent function in eq.~\eqref{eq:phi_p2} is renormalized by writing:
    \begin{align}
\phi(t,M;d)=\phi^{ren}(t,M,\mu;d)-\frac{t\lambda(\mu)}{6F_{\pi}^2} +\mathcal{O}(p^4)\,,
    \end{align}
    and the complete expression for pion electromagnetic form factor has the form:
    \begin{align}
    F_V^{\pi}(t)=1+2\phi^{ren}(t,M_\pi,\mu;d)+\phi^{ren}(t,M_K,\mu;d)+\frac{2 t L^r_9(\mu)}{F_\pi^2}+\mathcal{O}(p^4)\,.
    \end{align}
\par
   Other examples can also be found in the book of Donoghue, Golowich and Holstein~\cite{Donoghue:1992dd}.

    An important attribute of the external source technique, which was not
    explicitly available in the heuristic proposal of Weinberg is the ability to also
    account for the electromagnetic and weak interactions.  This promotes the framework of chiral perturbation theory to an effective theory of the
    Standard Model (that includes electromagnetic and weak interactions) and not just that of the strong interaction sector with pions, kaons and the $\eta$.

\section{ Extensions of chiral perturbation theory}\label{ECT}
While the previous section treated `standard’ ChPT and its development, there are various extensions that are needed when specific processes are to be investigated where the methods of the previous section are not sufficient. We present here a short overview only, for a deeper treatment, the references given should be consulted.    
 \subsection{The \texorpdfstring{$\eta'$}{}}
Without the axial anomaly, see eq.~\eqref{eq:eq2}, there would be 9 light mesons. Because
of it, the singlet axial current is not conserved, and the ninth meson becomes
massive; indeed, the $\eta'$ has a mass of 957 GeV, comparable to that of the
nucleons. In the (hypothetical) limit $N_C \rightarrow  \infty$ the
$\eta'$ is indeed massless, but $N_C = 3$, and the $\eta'$ is heavy. However, despite the large mass difference of the $8$ pseudoscalar mesons and the $\eta'$, its influence on many processes is substantial, and therefore, it must
be included in a systematic treatment. A successful way is to 
incorporate the $\eta'$ in the $U$-matrix, that is, taking a three-dimensional {\it{unitary}} matrix~\cite{Gasser:1984gg} of form: 
\begin{align}
    U(x)=e^{\frac{1}{3}i \phi_0(x)}e^{i \phi(x)}
\end{align}

and add a mass term for the $\eta'$. In the absence of mixing, $\phi_0(x)$ corresponds to $\eta'$. Furthermore, as the $\eta'$ and the external field $\theta$ are both $SU(3)$ singlet pseudoscalars, they transform (up to a sign) in the same way under chiral transformations, and therefore the sum $\phi_0 + \theta$ is invariant. This
means that everywhere in the Lagrangian where there are constants, they should be replaced by arbitrary functions of $\phi_0 + \theta$. This
brings of course new uncertainties. Nevertheless, as shown in ref.~\cite{Gasser:1984gg}, it is possible to draw some concrete conclusions,
in particular about the mixing of the neutral pseudoscalar mesons $\pi^0$, $\eta$, and $\eta'$. Furthermore, as can be seen from 
eq.~\eqref{eq:L4}, low energy constant $L_7$ can be modeled by the exchange of a chiral singlet pseudoscalar. In fact, the $\eta'$
does a good job. There are possibly further applications of this way to include the $\eta'$
and more details can be found in Kubis et al.~\cite{Gan:2020aco, Kubis:2022yox}.
As mentioned, in the large $N_C$ limit, the models show new interesting aspects. See refs.~\cite{Kaiser:1998ds,Kaiser:2000gs} for a
thorough investigation.

\subsection{Vector mesons}\label{ECT1}
The next heavier hadrons after the pseudoscalar mesons are the vector mesons, such as the $\rho$. Using the methods
used before, in particular, that means determining the correct transformation behavior of the vector mesons under the chiral symmetry, they
can be built into the chiral Lagrangian~\cite{Ecker:1988te}. This does not only contribute to processes with such vector meson,
but the vectors are also resonances that contribute to the low energy constants introduced in section \ref{sectionCL}.
In fact, as stated there, in many cases, they largely saturate the constants, thus giving a very successful model for them. Some recent articles on the subject are~\cite{Pich:2008xj, Jiang:2015dba, Portoles:2010yt}.

\subsection{Baryons} \label{ECT2}
The method can also be extended to include the baryon degrees of freedom see review ref.~\cite{Scherer:2009bt, Mai:2022eur}. In the manifestly Lorentz covariant framework, a problem arises because one
    cannot have a strict power-counting scheme.  On the other hand, inspired by
    the heavy quark effective theory, a heavy baryon version is available due to
    Jenkins and Manohar~\cite{Jenkins:1990jv}. A reformulated Lorentz invariant method via infra-red regularization due to Ellis~\cite{Ellis:1999jt,Ellis:1997kc}
     and Becher and Leutwyler~\cite{Becher:1999he}, along with  other versions such as extended on-mass-shell
    renormalization methods that have several advantages were proposed in ref.~\cite{Fuchs:2003qc}. The pion-nucleon $\sigma-$term (see ref.~\cite{Alarcon:2021dlz} for latest review) obtained from the baryon ChPT has also been useful for beyond the standard model physics considerations, especially studies related to the dark matter searches~\cite{Crivellin:2013ipa}.

     \section{Two and Three Body Rescattering}
    \label{TBS}
    
    Scattering processes provide important clues to the physics behind them. Furthermore, scattering is often  part of other processes, like decays, where the total amplitude also depends on the (re) scattering of the decay products. For instance, in a decay $K$ into two pions, the pions rescatter strongly, thereby (in some cases) influencing decisively the measured decay rate. 
    This is important in cases where the straightforward application of ChPT is not sufficient to explain the experimental results and must be supplemented by additional methods, such as unitarity conditions which sum up certain higher order corrections. We
    also note that there is a vast literature on scattering.
    
    The formalism presented in section~\ref{sectionCL} for the one-loop ChPT can be used to calculate the scattering amplitude involving the pseudoscalar Goldstone bosons. In the limit of isospin conservation, it is customary to introduce amplitudes of definite isospin in the s-channel $T^I(s,t)$, which may be related to specific physical charged states and depend on the process of $\pi \pi$ or $\pi K$. These isospin amplitudes can further be decomposed into partial waves as:
\begin{align}
T^I(s,t) = 32\pi\sum_l (2l+1)t^I_l(s) P_l( cos (\theta))\,.
\end{align}
where $t_l^I(s)$ is partial wave amplitude, $\theta$ is the scattering angle in the center of mass frame, and the $P_l$ are the Legendre polynomials. The $t^I_l(s)$ are complex above the threshold and are related by unitarity. For $\pi\pi$ scattering it has the following form~\cite{Gasser:1983yg}:
\begin{align}
    t^I_l(s)= \left(\frac{s}{s-4 m_\pi^2}\right)^{1/2}\frac{1}{2i}\left(\eta^I_l(s) e^{2i \delta^I_l(s)}-1\right)\,,
\end{align}
with $\delta^I_l$ being the phase shift and $\eta^I_l$ the elasticity parameter. For $\pi K$ scattering, an analogous expression can be found in ref.~\cite{Ananthanarayan:2000cp}.
It may be recalled that scattering lengths are the lowest order shape parameters appearing in the expansion of the real part of the partial wave amplitudes, and their expression near the threshold looks like this:
\begin{align}
\text{Re }( t^I_l(s)) =(q^2)^l (a^I_l + b^I_l q^2 +\mathcal{O}(q^4)), 
\end{align}
where $q^2$ is the square of the momentum transfer in the center of mass frame and $(q^2)^l$ denotes the centrifugal barrier. 
The scattering length of the lowest waves dominates the physical cross-section at low energies. The scattering lengths are also one of the important quantities for the pionium as the decay rate that is sensitive to $\vert a_0^0-a_0^2\vert^2$\cite{Colangelo:2001df}. DIRAC~\cite{DIRAC:2005hsg} and NA48 experiments ref.~\cite{NA482:2005wht,Batley:2009ubw} at CERN were aimed to measure the
S-wave scattering length difference in the $I=0$ and $I=2$ isospin channels and the observed values were in agreement with ChPT predictions in ref.~\cite{Colangelo:2001df}. Since pions and kaons are short-lived, one cannot do fixed target experiments, and the obtained scattering lengths are based on phase shift analyses.  Whereas $e^+e^- \to \pi^+ \pi^-$ is well studied experimentally and is related to the $I=1$ $P-$wave via the Watson theorem, the other phase shifts are well measured only at higher energies from $\pi N$ scattering.  At low energies, they are related to the form factors of $K_{l4}$ decays and have to be extracted using dispersion relations. In pion scattering, the suitable framework is dispersion relations with two subtractions which suffice due to the Froissart bound, which allows one to write a system of partial wave equations that leaves the two S-wave scattering lengths undetermined parameters. These Roy equations and the corresponding Roy-Steiner equations for $\pi K$ scattering have been studied for over 50 years.
They also provide a very useful framework for relating the dispersion relations to the chiral amplitudes, as shown in this section.\par

    The  $\mathcal{O}(p^6)$ behavior of these contributions can be calculated following the work of Bijnens et al.~\cite{Bijnens:1995yn} and Colangelo, Gasser and Leutwyler~\cite{Colangelo:2001df} and the $\pi\pi$ scattering amplitude to can be decomposed in the following form: 
    \begin{align}
        t^{I}_{\ell}(s)=t^{I}_{\ell}(s)_2+t^{I}_{\ell}(s)_4+t^{I}_{\ell}(s)_6+\mathcal{O}(p^8)\,.
    \end{align}
    At the leading order, the non-zero contributions from the S- and P- waves are given by:
    \begin{align}
        t^0_0(s)_2=\frac{2s-M_{\pi}^2}{32\pi F_{\pi}^2},\quad t^1_1(s)_2=\frac{s-4M_{\pi}^2}{96\pi F_{\pi}^2},\quad t^2_0(s)_2=-\frac{s-2M_{\pi}^2}{32\pi F_{\pi}^2}\,.
    \end{align}
   To $\mathcal{O}{(p^6)}$ accuracy, the imaginary parts of the $\pi\pi$ (and $\pi K$) scattering amplitude receive contributions only from the $S-$ and $P-$ partial waves and can be written in terms of the three functions of only one variable as:
    \begin{align}
        A(s,t,u)=C(s,t,u)+32\pi \bigg\lbrace&\frac{1}{3}U^0(s)+\frac{3}{2}(s-u)U^1(t)+\frac{3}{2}U^1(u)\nonumber\\&+\frac{1}{2}\left(U^2(t)+U^2(u)-U^2(s)\right)\bigg\rbrace\,,
    \end{align}
    where the first term must obey
    crossing symmetric and has the form:
    \begin{align}
        C(s,t,u)=c_1+s\hspace{.2mm} c_2+s^2 c_3+(t-u)^2c_4+s^3 c_5+s(t-u)^2 c_6\,.
    \end{align}
    It may be borne in mind that the real parts obtain contributions from the $\mathcal{O}(p^4)$ from higher waves as well. The $c_i$ are the subtraction constants of $U^i(x)$, which are also termed ``unitarity corrections". For $s-$channel with isospin $I=0,1,2$ have dispersion relation given by:
    \begin{align} 
        U^{0}(s) &=\frac{s^{4}}{\pi} \int_{4 M_{\pi}^{2}}^{\infty} d s^{\prime} \frac{\sigma\left(s^{\prime}\right) t_{0}^{0}\left(s^{\prime}\right)_{2}\lbrace t_{0}^{0}\left(s^{\prime}\right)_{2}+2 \operatorname{Re} t_{0}^{0}\left(s^{\prime}\right)_{4}\rbrace}{s^{\prime 4}\left(s^{\prime}-s\right)} \\ 
        U^{1}(s) &=\frac{s^{3}}{\pi} \int_{4 M_{\pi}^{2}}^{\infty} d s^{\prime} \frac{\sigma\left(s^{\prime}\right) t_{1}^{1}\left(s^{\prime}\right)_{2}\lbrace t_{1}^{1}\left(s^{\prime}\right)_{2}+2 \operatorname{Re} t_{1}^{1}\left(s^{\prime}\right)_{4}\rbrace}{s^{\prime 3}\left(s^{\prime}-4 M_{\pi}^{2}\right)\left(s^{\prime}-s\right)}\,, \\ U^{2}(s) &=\frac{s^{4}}{\pi} \int_{4 M_{\pi}^{2}}^{\infty} d s^{\prime} \frac{\sigma\left(s^{\prime}\right) t_{0}^{2}\left(s^{\prime}\right)_{2}\lbrace t_{0}^{2}\left(s^{\prime}\right)_{2}+2 \operatorname{Re} t_{0}^{2}\left(s^{\prime}\right)_{4}\rbrace}{s^{\prime 4}\left(s^{\prime}-s\right)}\,,
    \end{align}
     more details about various quantities appearing in this equation can be found in ref.~\cite{Colangelo:2001df}. \par
     The case when there is no three-channel crossing symmetry, and with unequal mass scattering is also accessible using a combination of fixed-t and hyperbolic dispersion relations, which were known in the literature after being suitably modified to account for
chiral counting, in order to saturate the dispersion relations using the imaginary parts
of the relevant S- and P- waves. 

    In the case of $\pi K$ scattering, the structure was analyzed by Ananthanarayan and B\"uttiker~\cite{Ananthanarayan:2000cp}
    and by B\"uttiker, Descotes-Genon and Moussallam~\cite{Buettiker:2003pp}. The $\pi K$ scattering amplitude to one loop can be decomposed into partial waves. Once one isospin channel amplitude is known, others or a combination of them can be obtained using the crossing symmetry relations. Like $\pi\pi$ scattering, these amplitudes can also be written in terms of functions of one variable as:
    \begin{align}
        T^+(s,t,u)&=Z_t^+(t)+Z^+_0(s)+Z^+_0(u)+(t-s+\frac{\Delta^2}{u})Z^+_1(u)+(t-u+\frac{\Delta^2}{s})Z^+_1(s)\\
        T^-(s,t,u)&=Z_t^-(t)+Z^-_0(s)-Z^-_0(u)+(t-s+\frac{\Delta^2}{u})Z^-_1(u)-(t-u+\frac{\Delta^2}{s})Z^-_1(s)\,.
    \end{align}
    The imaginary parts of the $Z$'s can be written in terms of the lowest partial waves as:
    \begin{align}
        \text{Im}\hspace{1mm}Z^{\pm}_0(s)&=16\pi \text{Im} f_0^{\pm}(s)\,,\\
        \text{Im}\hspace{1mm}Z^{\pm}_1(s)&=\frac{12\pi}{q^2_s} \text{Im} f_1^{\pm}(s)\,,\\
        \text{Im}\hspace{1mm}Z^{+}_t(s)&=\frac{16\pi}{\sqrt{3}} \text{Im} f_0^{I_{t}=1}(t)\,,\\
        \text{Im}\hspace{1mm}Z^{-}_t(s)&=6\sqrt{2}\pi \text{Im} \frac{f_0^{I_{t}=1}(t)}{p_tq_t}\,.
    \end{align}
    The details of various quantities appearing in these equations can be found in ref.~\cite{Ananthanarayan:2000cp}.\par 
    There are processes where it is necessary to account also for 3-particle rescattering, which is considerably more complicated. This is, for instance, the case for decays where phase space is limited. The best-known example is the decay of $\eta\to 3 \pi$ with significant data available for the cases of exclusively neutral, as well as neutral, and charged pions, in terms of the Dalitz plot as well as in terms of rates.  This rate is sensitive to the $u-d$ mass difference and, therefore of special importance in the determination of the quark mass ratio (Q)~\cite{Lanz:2011foc}. The original work of Khuri-Treiman~\cite{Khuri:1960zz} is based on the dispersive approach to study the final state interactions in $K\rightarrow3\pi$, and a set of integral equations are obtained and later to $\eta \rightarrow3\pi$ by Kambor, Wisendanger and Wyler~\cite{Kambor:1995yc} and Leutwyler and Anisowich~\cite{Anisovich:1996tx}.  The presence of final state interactions between the pion generates the branch cut in the amplitudes that starts from $4m_\pi^2$ in $s-$, $t-$, and $u-$ channels. As the centrifugal barrier suppresses the higher partial waves, the amplitude has a resemblance with the 2 body scattering where higher waves also start contributing from $\mathcal{O}(p^8)$. 
    The important difference between the two is that the three-body scattering also involves angular averages, which are difficult to perform. This difficulty has recently been overcome by an efficient method provided by Gasser and Rusetsky~\cite{Gasser:2018qtg}.
    \par 
    The scattering amplitude for $\eta\rightarrow3\pi$ can be decomposed into the contributions from isospin channel $I=0,1,2$ represented by $M_0$, $M_1$, $M_2$, which are the functions of one variable in Mandelstam variables. Following the detailed analysis of refs.~\cite{Anisovich:1966,Roy:1971tc,Anisovich:1993kn,Anisovich:1996tx}, the discontinuity in the amplitude has the form:
    \begin{align}
    \text{disc}M_I(s)=\theta(s-4M_\pi^2) \big\lbrace M_I(s)+\hat{M}_I(s)\big\rbrace \sin(\delta_I(s))e^{-i \delta_I(s)}
    \end{align}
    The first term in the braces receives contributions from the interactions of the $s$ channel, and the second term accounts for those coming from the $t$ and $u$ channels. The $\delta_I(s)$ are the phase shifts of the $\pi\pi$ scattering from the leading partial waves. The $t$ and $u$ channel contributions are given in terms of the angular averages of the $M_I$'s as follows:
     \begin{align}
        \hat{M}_0(s)&=\frac{2}{3}\langle M_0\rangle+2(s-s_0)\langle M_1\rangle+\frac{2}{3}\kappa\langle z M_1\rangle+\frac{20}{9}\langle M_2\rangle\\
        \hat{M}_1(s)&=\kappa^{-1}\big\lbrace3\langle z M_0\rangle+\frac{9}{2}(s-s_0)\langle z M_1\rangle-5\langle z M_2\rangle+\frac{3}{2}\kappa\langle z^2 M_1\rangle\big\rbrace\\
        \hat{M}_2(s)&=\langle M_0\rangle-\frac{3}{2}(s-s_0)\langle M_1\rangle-\frac{1}{2}\kappa\langle z M_1\rangle+\frac{1}{3}\langle M_2\rangle\,,
    \end{align}
    where 
    \begin{align}
        s_0&=\frac{1}{3}M_{\eta}^2+M_\pi^2\,\\
        \kappa(s)&=\sqrt{1-\frac{4M_\pi^2}{s}}\sqrt{(M_{\eta}^2-M_\pi)^2-s}\sqrt{(M_{\eta}^2+M_\pi)^2-s}\,\\
        \langle z^n M_I\rangle(s)&=\frac{1}{2}\int_{-1}^1dz z^n M_I(\frac{3}{2}s_0-\frac{1}{2}s+\frac{1}{2}z\kappa(s))
    \end{align}
    with $I=0,1,2$ and $n=0,1,2$. For more details, we refer to ref.~\cite{Colangelo:2018jxw}. \par 
     The details of the higher order corrections to $\mathcal{O}(p^6)$ for the three body decay of the  $\eta\rightarrow3\pi$ can be found in refs.~\cite{Gasser:1984pr,Bijnens:2002qy,Bijnens:2007pr}. The dispersive construction of amplitude can be found in Kampf et al.~\cite{Kampf:2011wr,Kampf:2019bkf} and small electromagnetic corrections to this process in Ditsche, Kubis and Mei\ss ner~\cite{Ditsche:2008cq}. A detailed analysis using Dalitz plot and modified non-relativistic effective field-theory in by Schneider, Kubis and Ditsche in ref.~\cite{Schneider:2010hs}. The determination of quark mass ratio from these decays are presented in  refs.~\cite{Lanz:2011foc,Colangelo:2011zz,Lanz:2013ku,Colangelo:2016jmc}. Cusps in $K\rightarrow3\pi$, which are relevant for the precise determination of the pion scattering lengths, are studied in ref.~\cite{Colangelo:2006va,Gasser:2011ju}, in $\eta \rightarrow3\pi$, effects of mixing of $\eta \eta'$ in the $\eta\rightarrow3\pi$ Leutwyler~\cite{Leutwyler:1996np}, dispersive analysis by Leutwyler and Anisovich in ref.~\cite{Anisovich:1996tx} and various topics related to three-body decays and dispersion relations are now covered in the book of Anisovich et al.~\cite{Anisovich:2013gha}.
    For a detailed review, we refer to refs.~\cite{Colangelo:2018jxw,Gan:2020aco}.    

\section{Generalized renormalization group and large chiral logarithms}
\label{GRG}

    In section~\ref{sectionCL}, we touched briefly upon infinities
    in the low energy constants, see eq.~\eqref{eq:eq12}.
    Such infinities are, of course, well-known in QED and other quantum field theories. 
    Historically such divergencies in the self-energy of an electron from classical electrodynamics led to the birth of quantum field theory. Schwinger, Tomanaga, Feynman, and Dyson gave a covariant description of QED which led to the consistent description to any order in the perturbation theory. The infinities are removed by redefinition in the bare parameters of the Lagrangian, a procedure termed renormalization. At that time, it was just a mathematical trick to tackle the divergences. The works of Stueckelberg and Petermann~\cite{StueckelbergdeBreidenbach:1952pwl}, Gell-Mann and Low~\cite{Gell-Mann:1954yli} showed that this procedure automatically incorporates the running of renormalized coupling constants, see eq.~\eqref{eq:eq13}. The renormalization group equations dictate the running and mixing of various operators with scales and have been used as a very useful technique that allows to sum up some of the large logarithmic corrections which are remnants of the renormalization procedure.

    Whereas the early discussion was mainly restricted to perturbation theory and renormalizable theories, Wilson~\cite{Wilson:1973jj,Wilson:1971bg,Wilson:1971dh} in the early 1970s further extended it to non-perturbative systems in order to understand critical phenomenons and gave a deeper insight to the physics at different scales. This has found numerous applications in various areas of physics ranging from condensed matter, statistical physics, and cosmology to particle physics. These ideas were later studied in great detail using the path integral by Polchinski~\cite{Polchinski:1983gv}. There are various approach to the renormalization group and we refer to refs.~\cite{Schwartz:2014sze,Peskin:1995ev,Hollowood,Jakovac:2016zkg,Gies:2006wv,Polonyi:2001se,Burgess:2020tbq,Baldazzi:2021lym}. An overview can be found in ref.~\cite{Huang:2013zaa}.

    While the concepts of renormalization are associated with renormalizable theories,
    one might ask how they work in non-renormalizable theories, such as ChPT, where the number of parameters increases to cancel the divergences appearing in the loop calculations. In particular, one may ask how to order the (large) leading logarithms (LL) which arise in the calculations.
    
    Li and Pagels~\cite{Li:1971vr} in the early seventies pointed out that a large logarithm of type $m^2_{\pi}\log(m^2_{\pi})$ appears in one-loop calculations involving pion loops. Weinberg calculated these logarithms in his famous paper on phenomenological Lagrangians~\cite{Weinberg:1978kz} using current algebra and the renormalization group for pion scattering. Later work of Gasser and Leutwyler~\cite{Gasser:1983yg} where systematic one-loop extension of ChPT was performed and significant $\sim25\%$ contribution at $1\GeV$ from such terms were obtained, especially the corrections to the lowest S-wave pion scattering length. The large logarithm contributions to the two-loop can be found in ref.~\cite{Colangelo:1995np} and have the following form:
    \begin{align}
        a^0_0=\frac{7m^2_{\pi}}{32F^2_{\pi}}\left(1-\frac{9}{2}\frac{m^2_{\pi}}{16\pi^2F^2_{\pi}}\log(m^2_{\pi}/\mu^2)+\frac{857}{42}\left(\frac{m^2_{\pi}}{16\pi^2F^2_{\pi}}\right)^2\log^2(m^2_{\pi}/\mu^2)\right)\,.
    \end{align} 
   
    The full two-loop contributions to scattering length read~\cite{Bijnens:1995yn}:
    \begin{align}
a_0^0=&\overbrace{0.156}^{\mbox{tree}}+\overbrace{0.039+0.005}^{\mbox{1~loop}}+\overbrace{0.013+0.003+0.001}^{\mbox{2~loops}}=\overbrace{0.217}^{\mbox{total}}\nonumber\\
 & \hspace{1.5cm} L \hspace{0.75cm}\mbox{anal.} \hspace{.8cm} k_i\hspace{1cm}L\hspace{0.8cm}\mbox{anal.}\nonumber\,,
    \end{align}
     \par 
    where $k_i$ are the contributions from the single as well as double chiral logarithms, which can be evaluated using the renormalization group~\cite{Colangelo:1995np,Weinberg:1978kz}. Bijnens, Colangelo and Ecker~\cite{Bijnens:1998yu,Bijnens:1999hw} extended the work on chiral double logarithms to the full meson sector. Clearly, logarithmic corrections can be large in ChPT, and a tool like the renormalization group would be useful.
    
    Indeed, Kazakov~\cite{Kazakov:1987jp} and Alvarez, Freedman, and Mukhi~\cite{Alvarez-Gaume:1981exa} discussed the extension of renormalization to arbitrary (non-renormalizable) theories in order to calculate leading and subleading divergences. These ideas were applied to ChPT by Buchler and Colangelo~\cite{Buchler:2003vw}, which required a new one-loop calculation at each order. The resummation of these large logarithms to all orders is still an open question in ChPT. However, the chiral logarithms have been of constant interest to understand many renormalizable and non-renormalizable theories as a toy model. Bissegger and  Fuhrer~\cite{Bissegger:2006ix} worked out a method to calculate the chiral logarithms for two flavors to any desired order in chiral limit using analyticity, crossing symmetry, and the Roy equations. They have also given the five-loop results for specific two-point scalar Green functions. Kivel, Polyakov, and Vladimirov~\cite{Kivel:2008mf} provided a method where a non-linear recurrence relation is obtained that efficiently calculates the leading logarithm (LL) to arbitrary loops for any non-renormalizable theories. This work was later extended for form factors in ref.~\cite{Kivel:2009az} and some results for the LLs for the massless $O(N + 1)/O(N)$ $\sigma$-model are also presented. This model, for $N = 3$, is equivalent to the chiral $SU(2) \times SU(2)$ model that describes the leading low-energy interaction of pions in the chiral limit. Later Koschinski, Polyakov, and Vladimirov~\cite{Koschinski:2010mr} provided a method to calculate the leading infrared logarithms to essentially unlimited loop order using only the tree-level results in the non-renormalizable massless effective theory and later to sigma models on an arbitrary Riemann manifold by Polyakov and Vladimirov in ref.~\cite{Polyakov:2010pt}. The LLs in the  massive case for a non-linear $O(N)$-sigma model are studied by Bijnens and Carloni in ref.~\cite{Bijnens:2009zi,Bijnens:2010xg} and extended to the anomalous sector by
    Bijnens, Kampf and Lanzin in ref.~\cite{Bijnens:2012hf}.  More recently, Ananthanarayan, Ghosh, Vladimirov, and Wyler~\cite{Ananthanarayan:2018kly} have generalized the massive case to arbitrary order in LL corrections for various $O(N)$ and $SU(N)$ models and a Mathematica code is provided that reproduces the existing results and calculates higher-order  results. Further development in two-dimensional effective field theories can be found in ref.~\cite{Polyakov:2018rdp,Linzen:2018pvj} and an extension to the baryon sector in ref.~\cite{Bijnens:2014ila}.
  \section{Weak interactions of pseudoscalar mesons}
  \label{Weak}
\begin{figure}[H]
        \centering
        \includegraphics[width=.5\linewidth]{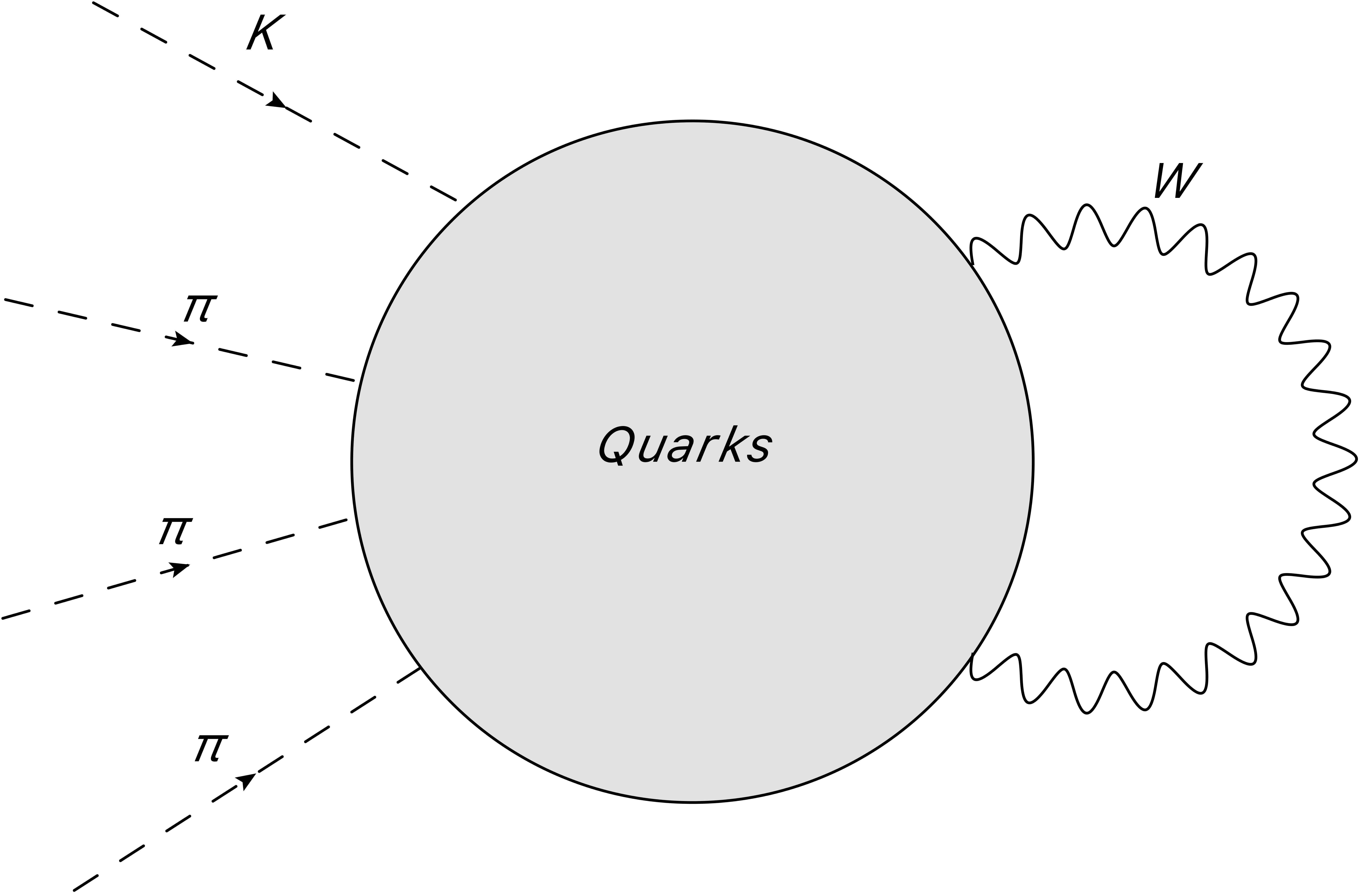}
        \caption{Illustration of the weak interaction of the pseudoscalar mesons. The large mass of the $W$ bosons is the reason why the external field method is appropriate~\cite{Binosi:2003yf}.}
        \label{fig:my_label}
    \end{figure}

    The ChPT formalism, especially when formulated with the external field method, is directly adaptable to weak processes involving the pseudoscalar mesons, such as the decays $K \to \pi \pi$, and others. A recent example illustrating the persistent importance of ChPT is the rare decays involving (hypothetical) new light particles such as axions~\cite{Bauer:2021wjo}. The systematic expansion in powers of momentum and quark masses allows analyzing seriously many `small' effects. An illustration of how the weak interactions fit into the external field method with well-defined transformation properties is given in figure~\ref{fig:my_label}. The large size of $M_W$ compared to the QCD scale of a few GeV makes it clear that any interaction of gluons that affect the $W$ bosons is tiny: It would involve the strong coupling constant at the $M_W$ scale and further suppression factors $1/M_W$.
    
    The basis for extending ChPT to the weak interactions was laid down in ref.~\cite{Kambor:1989tz}. It gives a systematic treatment of ChPT for weak interactions and extended the weak interactions Lagrangian to $\mathcal{O}\left(p^4\right)$. It is based on several previous works; here, we mention only the pioneering paper by Cronin~\cite{Cronin:1967jq}.
    
    To construct the weak chiral Lagrangian, we need the form of the external field that represents the weak interactions. The (chiral) symmetry properties of the weak interaction follow from the fact that they arise from the symmetric product of two left-handed charged octet currents:
    
    \begin{equation}
        \mathcal{L}_{\Delta S=1}=g\hspace{1mm}\{J^2_{1\mu},J^{1\mu}_3\}_+ + g^*\hspace{1mm}\{J^1_{2\mu},J^{3\mu}_1\}_+\,,
    \end{equation}
    with 
    \begin{equation}
        J^2_{1\mu}=J_{1\mu}+i\hspace{.2mm}J_{2\mu},\quad J^3_{1\mu}=J_{4\mu}+i\hspace{.2mm}J_{5\mu}
    \end{equation}

    and where the (numeral) indices refer to the position in the $3 3$ flavor matrix and $\{ , \}_+$ denotes the symmetric product. This implies that the
    weak interactions transform as $ (8)_L$ and $(27_L,\hspace{2mm})$. We note that the CP-invariant and the CP-odd parts can be conveniently separated in ref.~\cite{Kambor:1989tz}.
    
    Using now the expressions 
    \begin{equation}
        L_{\mu}=i\hspace{.2mm}U^+\nabla_{\mu}U
    \end{equation}
    for the left-handed meson currents, we can write the octet CP invariant effective weak operator as
    \begin{equation}
\mathcal{L}^{(8)}_{WI}=c_2\langle\lambda_6\nabla_{\mu}U^\dagger\nabla^{\mu}U\rangle=c_2 \langle\lambda_6 L_{\mu}L^{\mu}\rangle
    \end{equation}
    where the octet property is manifest in the matrix $\lambda_6$ \footnote{Since we consider $K-$decays, only the transition from an $s-$quark to a $d-$quark, that is only the Gell-Mann matrices with elements $(2,3)$ contribute}.  At $\mathcal{O}\left(p^2\right)$, also a second operator can
    be written as:
    \begin{equation}
        \mathcal{L}^{8'}_{WI}=c_5\langle\lambda_6\left(\chi^\dagger U +U^\dagger \chi\right)\rangle
        \label{eq:kmw237}
    \end{equation}
    The CP-invariant effective weak chiral Lagrangian transforming as $(27_L)$ is constructed from the octet components of $L_\mu$:
    \begin{equation}
        \mathcal{L}^{27}_{WI}=c_3\{3\langle\left(Q^2_3+Q^3_2\right)L_{\mu}\rangle\langle Q^1_1 L^{\mu}\rangle+2\langle Q^2_1 L_{\mu}\rangle \langle Q^1_3 L^{\mu}\rangle+2\langle Q^1_2 L_{\mu}\rangle \langle Q^3_1 L^{\mu}\rangle\}
    \end{equation}
    where the matrices $Q^i_j$ have a $1$ in the position ${i,j}$ and are zero otherwise. We note that there is only one operator in this case. An application of this CP-invariant operator to $K\rightarrow\pi \ell\overline{\ell}$ process at one-loop can be found in ref.~\cite{Ananthanarayan:2012hu}.
    
    As to the $CP$-violating Lagrangian, it is obtained from the above by replacing $\lambda_6$ by $\lambda_7$ and appropriate changes
    in the operators transforming as $27-$plet
    \begin{align}
        \mathcal{L}^{-}_{WI}=&c_2^-\langle\lambda_7L_{\mu}L^{\mu}\rangle+c_5^-\langle\lambda_7\left(\chi^{\dagger}U+U^{\dagger}\chi\right)\rangle\nonumber\\&+c_3^-\{3\langle\lambda_7L_{\mu}\rangle\langle Q^1_1 L^{\mu}\rangle+2\hspace{.2mm}i\hspace{.2mm}\left(\langle Q_1^2L_{\mu}\rangle\langle Q_3^1 L^{\mu}\rangle-\langle Q_2^1L_{\mu}\rangle\langle Q_1^3 L^{\mu}\rangle    \right)  \}
    \end{align}
    
    We note that the $\Delta S=2$ operator is required for the calculation of the mass difference in the $K^0$-$\bar{K}^0$ mixing, which
    transforms as a $27$-plet, and it is obtained by setting
    the tensor components to their appropriate values  (see ref.~\cite{Kambor:1989tz}).
    
    It is well known that the second octet operator in eq.~\eqref{eq:kmw237} does not contribute to physical processes. The operator is, in fact, proportional
    to the variation under a suitably chosen symmetry and thus to a divergence of a conserved (Noether) current. Since the operator does
    not carry momentum, the matrix element vanishes. In ~\cite{Kambor:1989tz}, the argument is extended to the one-loop level. We note
    here, however, that in processes where the scalar external field is not just $\chi$, but variable, this statement might not hold.
    
    While in ref.~\cite{Kambor:1989tz}, a complete basis of the weak operators at  $\mathcal{O}\left(p^4\right)$ is given, subsequent analyses showed that the basis could be further
    reduced, see ref.~\cite{Ecker:1992de} and ref.~\cite{Bijnens:2002vr}. 
    The complete $\mathcal{O}\left(p^4\right)$ Lagrangian containing 37 operators can be found in those papers. We also note that not all of these contribute to the decay of kaons into pions which make the calculations simpler and the predictions better.
    
   Much like in the strong interaction case discussed before, the application of the  $\mathcal{O}\left(p^4\right)$ Lagrangian to physical processes is used to determine the coupling strengths of the low energy operators, the LECs. In ref.~\cite{Bijnens:2002vr} the decay $K \to 3 \pi$ is analyzed. The order $\mathcal{O}\left(p^4\right)$ gets contributions from the operators mentioned and loop diagrams whose vertices are those of the lowest order interactions. For an improved treatment, see ref.~\cite{Cirigliano:2011ny}. Also, kaon decays are again considered as a laboratory for rare processes and recent progress can be found in ref.~\cite{NA62KLEVER:2022nea}.
   \section{Selected applications of Chiral Perturbative Theory.}
\label{APP}
As already mentioned, ChPT has numerous applications in describing low-energy processes. In some cases, the
precision reached is very high and allows for testing fundamental physics. Here we review but a few such
cases. \par As mentioned before in section \ref{sectionCL}, the masses of the quarks can be determined quite precisely using the chiral formalism from adequate phenomenological studies, such as of the $\eta\rightarrow3\pi$ decay (see section \ref{TBS}). Input from the lattice and QCD sum rules increases the accuracy. These studies have confirmed that the up quark mass $m_u$ is  non-zero~\cite{Aoki:2021kgd}.

The predominant decay of $\pi^0$ into the two photons proceeds via the chiral anomaly; the prediction for the rate is $\Gamma\left(\pi^0\to\gamma\gamma\right)=7.760\mathrm{ \hspace{1 mm} eV}$, in remarkable agreement  with $\Gamma\left(\pi^0\to\gamma\gamma\right)=7.82\pm0.14\mathrm{(stat.)}\pm17 \mathrm{(syst.)}$ eV  obtained from the high precision experimental finding of  PrimEx-II~\cite{Larin:2020bhc} experiment.\par 
Other processes such as $\pi\pi$ and  $\pi K$ scattering require detailed analysis using $SU(2)$ and $SU(3)$ versions of the ChPT. The scattering amplitude of these processes, when expanded in terms of the partial amplitudes, results in the notion of the scattering lengths, and their experimental inputs can be used to fix some of the low-energy constants. An explicit expressions $\pi\pi$, $\pi K$, and $K K$ scattering lengths to $\mathcal{O}\left(p^4\right)$ can be found in ref.~\cite{Aoki:2021kgd} and references therein. Interestingly, for the $\pi\pi$ interaction, the scattering lengths for the $I=0$ isospin channel have a positive sign and are larger than 3.5 times in magnitude compared to the $I=2$ isospin channel, which has a negative sign. These signs correspond to the repulsive and attractive nature of the interactions in these channels. Furthermore, the phase shift analysis of the $\pi\pi$ scattering has been found to be a very useful ingredient in quantifying the hadronic contributions to the anomalous magnetic moment of the muon (see below). Readers can find further details on the form factors in  refs.~\cite{Ananthanarayan:2022wsl,CGL,CS,Blum:2021fcp,Andersen:2018mau} and references therein. \par
Of particular interest is the anomalous magnetic moment of the muon. It is one of the testing grounds for the standard model and has been the topic of constant interest in the particle physics community~\cite{Jegerlehner:2017gek,JegNyf}. The results from the Brookhaven National Laboratory (BNL) found tension with the predictions of the standard model a little over 3$\sigma$ in ref.~\cite{BNL}. Further development in both the theory and experiment side has taken place and is summarized in ref.~\cite{WP}. The most recent experiment in Fermilab aimed to study this issue with improved purity of the beam and detector components and found agreement with the results of BNL with a smaller central value. Their combined results have has now established the discrepancy at $~4.2\sigma$. These results can be found in a set of publications in refs.~\cite{FL1,FL2,FL3,FL4}. The main source of the discrepancy comes from the hadronic vacuum polarization contributions and another somewhat less numerically important but relatively larger uncertainty known as the hadronic light by light scattering contributions. An excellent summary of all these discussions was recently presented, see slides of~\cite{Gilberto}, and for details,  we refer to  ref.~\cite{Colangelo:2022jxc,Borsanyi:2020mff,ACD,Ananthanarayan:2022wsl} and references therein. Some of these hadronic light-by-light contributions, as well as those contributions to $(g-2)_{\mu}$ involve related processes where transition form factors play an important role. These form factors are the complex functions obeying the unitarity and analyticity conditions, which dictate their behavior in the complex plane. However, their values for a given kinematical region can be fixed by the available information from the experiments or lattice simulations. In some cases, the Watson theorem relates the phase shift of the scattering amplitude to the phase of the form factor. One of them that is worth mentioning is the transition form factor for the  $\omega \pi^0$ for which discrepancies between experimental data and results from dispersion theory were reported for low energy region; see ref.~\cite{Ananthanarayan:2016icw} and references therein for details. However, these discrepancies can be studied in a model-independent way using the method of unitarity bounds\cite{Okubo:1971jf,Okubo:1971my} combined with the functional analysis method~\cite{Ananthanarayan:2014pta} to find the bounds on the $\omega\pi^0$ form factor. These functional methods have found numerous applications in hadron physics and are now available in the form of a textbook in ref.~\cite{Caprini:2019osi}. Recently, some agreement between experimental data with new analysis based on subtracted Khuri-Treiman equations has been reported for $\omega\pi$ transition form factor in ref.~\cite{JPAC:2020umo}.\par
    Of course, several other examples can be studied using chiral perturbation theory, and many of them can also be found in the supplementary Mathematica~\cite{Mathematica} notebooks of ref.~\cite{Ananthanarayan:2012dq} and references therein. The following publicly available codes are recommended to study some of the processes:
    \begin{itemize}
    \item Ampcalculator by Unterdorfer and Ecker~\cite{Unterdorfer:2005au}. 
    \item Phi by Orellana which calculates $\mathcal{O}\left(p^4\right)$ corrections to one loop and already included in FeynCalc 9.0~\cite{Shtabovenko:2016sxi} and later versions. 
    \item The Mathematica-based code to study the $\pi\pi$ scattering, and Scalar and Pseudoscalar Form Factor and new additions to meson-meson scattering using $U(3)-$ChPT can be found in the link~\cite{Oller:U3}.  
    \item Mathematica notebooks with many solved examples by Ananthanarayan, Das, and Imsong in ref.~\cite{Ananthanarayan:2012dq}
      \end{itemize}
   
    \section{Other Effective Theories for the strong interactions}
    \label{OSI} 
    While ChPT is designed for phenomena where momentum exchange is below $1 \GeV$, one must also deal with QCD at higher energy scales. There are several effective methods proposed and used in particle physics to account for the strong interactions, in particular for their leading effects. 
    They allow for adapted calculations in processes where strong interactions are important.
    With the huge harvest of ever-improving experimental data over the last decades, such methods are, in fact, necessary to explain and exploit these results as fully as possible. In particular, they are used to uncover a possible still more fundamental theory than the standard model.\par    
    Characteristic for these situations is the presence of (two) very different scales, $m_1 \ll m_2$, that are relevant for the processes considered. Then, typically, either an expansion in the small quantity $m_1/m_2$ is possible, or there are large logarithms of the form 
    $\log(m_1/m_2)$ originating in loops, see eq.~\eqref{eq:eq13} for details. \par    
    At present, the study the weak interactions and possibly other fundamental physics involves 
    three important energy scales: (1) The weak scale, $M_W$ is of the order of 100$\GeV$, (2) the mass scale of the heavy quarks $b$ and $c$ (several $\GeV$), and (3) the QCD scale 
    $\Lambda_{QCD}$ of about $\sim1/3\GeV$ where the confinement effects set in.\par
    At $M_W$, the strong coupling constant $\alpha_s$ is about $\sim 0.118$, and the strong interactions are perturbative (asymptotic freedom). For the heavy quark mass scale, $\alpha_s$ is about 0.25. This still allows for perturbative calculations, but their precision is limited. While for the $b$ quark mass, this treatment seems appropriate, the scale of the charm quark offers
    substantial difficulties. Even more involved is the situation for the strange quarks; that is the physics of kaons. We
    note, however, that because of the `Cabibbo suppression', the decays of the $b$ and $s$ are easier to study than those of the $c$ quarks. For recent and updated overviews, see refs.~\cite{Buras:2022irq, Buras:2022wpw} and references therein, or refs.~\cite{Buras:2021rdg,Cervenkov:2022lpm, Destefanis:2022esu} for charm.
    
   We note that the methods to be described are primarily used to analyze and calculate the effects that the strong interactions have on investigations of fundamental parameters and theories, such as the elements of the Cabibbo-Kobayashi-Maskawa (CKM) matrix. Of course, there are still properties of the strong interactions themselves and it is interesting to understand them, for instance, the spectrum and decay width of the charm quark systems. Recent progress in this sector can be found in ref.~\cite{Kato:2018ijx,Chen:2022asf}.
    
    \subsection {Extended Effective weak Theory}
\begin{figure}[H]
        \centering
        \includegraphics[width=.4\linewidth]{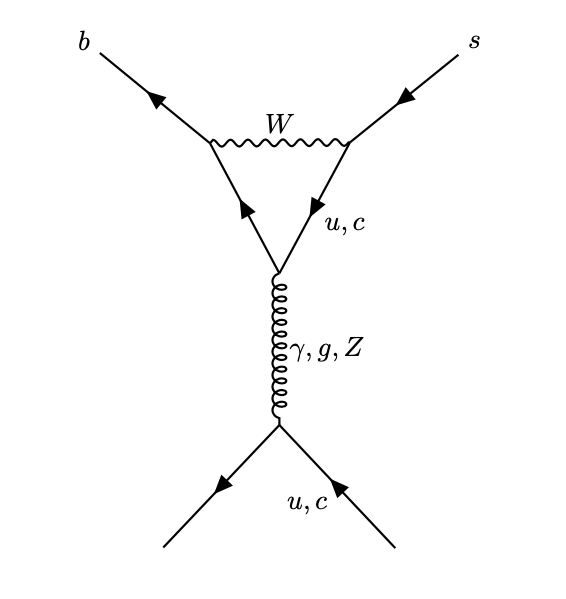}
        \caption{Penguin diagram contributing to $B\rightarrow X_s \gamma$.}
        \label{figpenguin}
    \end{figure}

    This methodology was put forward after the discovery of asymptotic freedom and the realization that QCD, in fact, allows for perturbative calculations. It is used mainly to investigate weak interaction processes of the heavy quarks $b$ and $c$, but also the (weak) decays of the $s$ where it was first applied. It is an extension of the original 4-Fermi theory and allows
    to include loops of the electroweak and strong interactions in a systematic way. In particular,
    the strong interaction effects can be calculated reliably in the interval between the weak scale  $M_W$ and the mass of the heavy quarks, thereby taking into account the large logarithms.
    Work on this began in the mid-seventies. Shifman, Altarelli, Cabibbo, Maiani, Petronzio, Ellis, Gaillard, Lee, Gilman, Wise, and Buras are but a few that have made important contributions and perfected the theory. For some of the original literature, see the refs.~\cite{Altarelli:1974exa, Gaillard:1974nj, Gilman:1978wm}.
    We will give only a rudimentary introduction for many details of this advanced, by now standard subject; see the book by Buras~\cite{Buras:2020xsm}, which offers an in-depth and updated treatment; and for an even more recent update, we refer to ref.~\cite{Buras:2022irq,Albrecht:2021tul}.
    The basic idea is that at energies below $M_W$, the dynamical fields are the quarks (except the 
    top quark), gluons, and photons (or other light, undiscovered particles). Thus the weak Hamiltonian operator $\hat{\mathcal{O}}$ can be written as a series of operators consisting of the quark fields of interest, gluons, and photons with increasing powers of $1/M_W$; in reality, the important power is $1/M_W^2$. These operators must satisfy the symmetries required for the process at hand and are usually ordered according to increasing orders of $1/M_{W}^2$ \footnote{In cases where the top quark is important, there are also inverse powers of top quark mass}. For consistency, all operators that can contribute to the process at the desired order in the strong and electromagnetic coupling constant must be considered. This implies that not only the original left-left Four-Fermi operator ($W$-exchange) is present, but several others are generated through loop corrections. A famous example is the so-called penguin operator (See Fig \ref{figpenguin}).

 For instance, the operator for the decay $B \rightarrow X_s \gamma$ ($X_s$ denotes an inclusive hadronic state with the total strangeness of one) takes the form:
    \begin{align}
\mathcal{H}_{\mathrm{eff}}(b \rightarrow s \gamma)=-\frac{4 G_{\mathrm{F}}}{\sqrt{2}} V_{t s}^{*} V_{t b}\left[\sum_{i=1}^{6} C_{i}\left(\mu_{b}\right) Q_{i}+C_{7 \gamma}\left(\mu_{b}\right) Q_{7 \gamma}+C_{8 G}\left(\mu_{b}\right) Q_{8 G}\right]\,,
\end{align} where the `magnetic' penguin operators in the above are given by:
\begin{align}
Q_{7 \gamma}=\frac{e}{16 \pi^{2}} m_{b} \bar{s}_{\alpha} \sigma^{\mu v} P_{R} b_{\alpha} F_{\mu \nu}, \quad Q_{8 G}=\frac{g_{s}}{16 \pi^{2}} m_{b} \bar{s}_{\alpha} \sigma^{\mu v} P_{R} t_{\alpha \beta}^{a} b_{\beta} G_{\mu v}^{a}\,.
\end{align}
More details about these equations can be found in the book of Buras~\cite{Buras:2020xsm}.
  Here, the operators $Q_{1}... Q_{6}$ are four-Fermi operators. There are six instead of only one because gluon exchanges rearrange
 the color order\footnote{this is indeed the crucial point of using an effective theory in that {\it{all}} operators consistent with the symmetries must be included.}. The coupling constants (as well as the quark field operators) depend on the 
    scale $\mu$ (see eq.~\eqref{eq:eq13}). The relevant scale for the decay at hand is of the order of $m_b$. On the other hand, the constants of the effective Hamiltonian can be perturbatively calculated at the high scale, $M_W$. Because of (weak and electromagnetic) loops, there can be more contributing operators beyond the simple $4-$Fermi interaction at the scale $m_b$.  To connect the two scales, the renormalization group is employed. This leads to a systematic expansion in the strong and electromagnetic coupling constants and the summing up of the large logarithms $\log(m_b/m_W)$. This procedure has led to a (almost) complete understanding of the weak parameters (such as the parameters of the Cabibbo-Kobayashi-Maskawa matrix) and, in particular, an understanding of $CP$ violation. 
    The status of deviations from the standard model in the heavy flavor sector can be found in ref.~\cite{Albrecht:2021tul}.
    For a detailed description of the method and the 
    results obtained, see ref.~\cite{Buras:2020xsm}. Note that this method 
    best applies to inclusive hadronic decay products (that is why in
    the above case, the final state is $X_s$, rather than an exclusive state, such as $K \pi$).

   \subsection{Heavy Quark Effective Theory} 
   While the effective weak theory described above pertains to the energy interval $m_b - M_W$,
   the heavy quark effective theory, HQET, deals with scales below $m_b$ in processes involving $b$ quarks, such as the $B$-meson. Since the typical momenta inside a QCD bound state are of the order of the strong scale $\Lambda_{QCD}$, which is much smaller than $m_b$, the $b$ quark is only lightly `shaken' and can therefore be considered at rest in a first approximation. Therefore, for an arbitrary heavy quark $Q$, we write the momentum of the quark as:
   \begin{align}
       p^{\mu} = m_Q v^{\mu} + k^{\mu}
       \label{heavymom}
   \end{align}
   where $v$ is the four-velocity of the hadron containing the heavy quark, and $k$ is of the order of $\Lambda_{QCD}$, and thus much smaller than $m_Q$. This decomposition allows, similar to the well-known treatment in atomic physics, to divide the spinor 
   into a dominant `upper' component and a `lower' one which is suppressed by $1/m_Q$.
   Thus, the idea is to construct an effective theory in which the upper component($h_v(x)$) is dynamical, and the lower one($H_v(x)$) is integrated out. This can be achieved by suitable projections of the $Q$
   quark spinor~\cite{Buchalla:1995vs}:
   \begin{align}
\Psi(x)=e^{-i m_{Q} v \cdot x}\left[h_{v}(x)+H_{v}(x)\right] \,.
\end{align}
  The upper and lower component is obtained by the relation:
  \begin{align}
      h_v(x)&=e^{i m_Q v\cdot x}\frac{1+\slashed{v}}{2}\Psi(x)\,,\\
      H_v(x)&=e^{i m_Q v\cdot x}\frac{1-\slashed{v}}{2}\Psi(x)\,,      
  \end{align}
  and in the case of heavy antiquark, the substitution of $v\rightarrow -v$ is made.
  Indeed, for $1/m_Q\rightarrow0$, the small
   the component can be integrated out~\cite{Georgi:1990um,Eichten:1989zv,Mannel:1991mc} and the theory has an extra spin-symmetry. The leading order (in $1/m_Q$) Lagrangian has the form:
   \begin{align}
\mathcal{L}_{\text {eff }}=\bar{h}_{v} i v \cdot D h_{v}+\mathcal{L}_{\text {light }}
\end{align}
   and other terms involving the heavy quark field are rearranged as an expansion in $1/M_Q$, and the Lagrangian for light degrees of freedom (quarks and gluons) is given by:
   \begin{align}
       \mathcal{L}_{light}=-\frac{1}{4}\text{Tr}\left(G_{\mu\nu}G^{\mu\nu}\right)+\sum_q \Psi_q\left(i \slashed{D}-m_q\right)\Psi_q\,.
   \end{align}
   This formalism has been extensively used in the literature to extract the CKM elements ($\vert V_{cb}\vert$, $\vert V_{ub}\vert$, heavy flavor sum rules, and the description of heavy hadron decays. More details can be found in refs.~\cite{Bigi:1997fj,Neubert:1993mb,Mannel:2020ups,HFLAV:2019otj,Lenz:2014jha,Lenz:2022rbq}.

\subsection{NRQCD and pNRQCD}
The heavy quark expansion used above is not suitable to describe a meson with two heavy quarks (like charmonium or the $\Upsilon$). In HQET, the kinetic energy is a $1/m_Q$ effect and is taken as a perturbation. But for a bound state, it plays an important role in balancing the potential energy and, therefore, should be present at leading order. The necessary formalism was provided by Bodwin, Braaten and Lepage in ref.~\cite{Bodwin:1994jh} and is known as NRQCD. Such systems also have additional scales, such as relative momenta $p\simeq m v $(soft) and a kinetic energy, $E_k\simeq m v^2$(ultrasoft scale), constructed out of the mass of heavy quark ($M$) and its velocity ($v\sim\alpha_s<<1$). For the bottomonium system, $v^2\sim 0.1$ and for charmonium systems, $v^2\sim 0.3$. The hierarchy scales in the system are as follows:
  \begin{align}
      m_q(\text{hard})\gg m_Q v\gg m_Q v^2\,.
      \end{align}
 The Lagrangian is expressed as an expansion in $m_Q v/m_Q$ and ($m_Q v^2/m_Q$), and at leading order in $1/m_Q$, it has the following form:
  \begin{align}
      \mathcal{L}_{\text{NRQCD}}=\psi^\dagger \left(i D^0 +\frac{\vec{D}^2}{2M}\right)\psi+\chi^\dagger \left(i D^0 -\frac{\vec{D}^2}{2M}\right)\chi+\mathcal{L}_{\text{light}}
  \end{align}
  where $ iD^0=i \partial_0-g A^0$ and $\psi(\chi)$ is the Pauli spinor field of fermion (antifermion). It should be noted that the presence of the two dynamical soft and ultrasoft scales can complicate the calculations and interfere with the power-counting and the non-perturbative effects. The NRQCD is numerous applications in the threshold production of top-quark pairs in electron-positron annihilation, spectroscopy of heavy charmonium and bottomonium bound states~\cite{Pineda:2011dg}, determination of heavy quark masses, strong coupling constant, and in the understanding of the vacuum structure etc. A modified version of the NRQCD has been recently proposed in ref.~\cite{Biswal:2022miy,Biswal:2022eld} for the production of the $J/\ensuremath{\Psi}$, $\ensuremath{\Psi}'$, and $\chi_c$. For more details, we refer to ref.~\cite{Brambilla:2004jw,Brambilla:2019esw}.\par
  Another interesting system that can be constructed out of NRQCD is the potential NRQCD (pNRQCD)~\cite{Pineda:1997bj,Brambilla:1999qa,Brambilla:1999xf}. It is obtained by integrating out the soft degrees of freedom. The leading order in $1/m_Q$ and multipole expansion in $r$, the Lagrangian has the following form:
 \begin{align}
\mathcal{L}^0_{\mathrm{pNRQCD}}&=\operatorname{Tr}\left\lbrace\mathrm{S}^{\dagger}\left(i \partial_{0}-V^{(0)}_{s}(r)\right) \mathrm{S}+\mathrm{O}^{\dagger}\left(i D_{0}-V_{o}(r)\right) \mathrm{O}\right\rbrace -\frac{1}{4} F_{\mu \nu}^{a} F^{\mu \nu a}\nonumber
\end{align} 
  where $S$ and $O$ are the singlet and octet fields. The resulting EFT has a resemblance to the Sch\"odinger equation as  the matching coefficients $V_i(r)$ play the role of the potential between the heavy quark. The equation of motion for the singlet case is :
  \begin{align}
  i \partial_0 S=\left(\frac{\mathbf{p}^2}{m}-V^{(0)}_s(r)\right)S
  \end{align} 
and depending on which scale is closer to $\Lambda_{QCD}$, different versions of pNRQCD (strongly or weakly coupled) are used for quarkonium. When there is no other scale between the soft and ultrasoft scales known as weakly coupled pNRQCD, the leading order static potentials have the form:
  \begin{align}
V_s^{\left(0\right)}=-C_F \frac{\alpha_{V_s}(r)}{r},\quad V_o^{\left(0\right)}=\left(\frac{C_A}{2}-C_F\right) \frac{\alpha_{V_o}(r)}{r}
  \end{align}
and $V_{s/o}(r)$ has a perturbative expansion in the strong coupling constant. These potentials have now been computed numerically to there-loop in refs.~\cite{Smirnov:2009fh,Anzai:2009tm} and analytically in ref.~\cite{Lee:2016cgz}. Some ultrasoft contributions to static energy in the weak coupling limit are already known to $\mathcal{O}(\alpha_s^4)$~\cite{Brambilla:2006wp} and two of us have given Pad\'e prediction for $\mathcal{O}\left(\alpha_s^4\right)$ term to $V_s(r)$ in ref.~\cite{Ananthanarayan:2020umo}. The QCD static potential has been a very useful quantity in the determination of the strong coupling constant $\alpha_s$ as it can be calculated to very good precision on the lattice~\cite{Bazavov:2019qoo}. Recent updates of $\alpha_s$ from static energy can be found in ref.~\cite{dEnterria:2022hzv,Komijani:2020kst,Ayala:2020odx} and references therein. There are several packages available in the literature that can be used to study non-relativistic systems. Recently, Brambilla et al.~\cite{Brambilla:2020fla} have published the publicly available Mathematica-based package FeynOnium that can be used to study the NREFTs to one loop. Another useful package relevant to studying the threshold quarkonium system is \texttt{QQbar\_threshold} by Beneke et al.~\cite{Beneke:2016kkb}. A detailed review on NRQCD, pNRQCD, and a description of quarkonia from these EFTs can be found in refs.~\cite{Brambilla:2004jw,Brambilla:2010cs}.
   \subsection{Heavy-light mesons}
  There exist some mesonic states with heavy and light quarks, and one may ask how to combine HQET and ChPT to study their production and decay. This issue has indeed been taken up by Burdmann and Donoghue~\cite{Burdman:1992gh}, Wise~\cite{Wise:1992hn} and Yan et al.~\cite{Yan:1992gz}, and is now known as heavy meson ChPT(HMChPT).
   It is formulated on the fact that the mass difference between the heavy meson and its excited state scales as $\sim 1/M_Q$, which can be of the order of a few $\MeV$s for heavy mesons such as $B$ meson. Heavy quark symmetry relates to the couplings of the $B$ and $B^*$, and it also relates to other mesons such as $D$ as long as the charm quark can be treated as heavy. A meson with one heavy quark can be labeled by the light quark spin $j_l$ and states with spin $j_l\pm\frac{1}{2}$ are degenerate due to heavy quark spin symmetry. Due to this fact, a consistent description of a heavy light system requires an excited state such as $B^*$ for $B$ systems, as their production will require much less energy than the pion mass. Since the energy involved are less than the pion mass, an extension to the chiral framework can be merged with the HQET.\par   
   Degenerate triplets of spin-zero mesons $P_a$ ($a=u,d,s$) and spin-one meson $P^*_a$ triplets are obtained by combining the spins of heavy and light quark spins using the heavy quark spin symmetry. These fields can be used to define the $4\times4$ matrix $H_a$, given by:
   \begin{align}
       H_a=\frac{\left(1+\slashed{v}\right)}{2}\left[P^*_{a\mu}\gamma^\mu-P_a\gamma_5\right]
   \end{align}
   where $P^*_{a\mu}$ is an operator that destroys a $P*_a$ meson with velocity $v$ and satisfies:
   \begin{align}
       v^\mu P^*_{a\mu}=0\,.
   \end{align}
   Defining $\overline{H}_a$ as:
   \begin{align}
    \overline{H}_a&\equiv\gamma^0 H^\dagger_a\gamma^0=
    \left[ P^{*\dagger}_{a\mu}\gamma^\mu+P^\dagger_a\gamma_5\right]\frac{\left(1+\slashed{v}\right)}{2}\,,
   \end{align}
   then most general leading order Lagrangian to describe the strong interaction between pseudo-Goldstone boson with heavy meson is given by:
   \begin{align}
       \mathcal{L}=&-i \text{Tr}\left(H v\cdot\partial \overline{H}\right)+\frac{F_\pi^2}{8}\text{Tr}\left\lbrace\partial^\mu U \partial_\mu U^\dagger\right\rbrace+\frac{i}{2} \text{Tr}\left(H v^\mu \left[U^\dagger\partial_\mu+U \partial_\mu U^\dagger\right]\overline{H}\right )\nonumber\\&+\frac{i g}{2} \text{Tr}\left( H \gamma_\nu \gamma_5 \left[U^\dagger\partial^\nu-U \partial^\nu U^\dagger\right]\overline{H}\right)-\frac{\Delta}{8}\text{Tr}\left(H \sigma^{\mu\nu}\overline{H}\sigma_{\mu\nu}\right)+\dots \,,
       \label{eq:HL1}
   \end{align}
   where $\Delta=m_{P^*}-m_{P}$, g is the axial coupling constant and field $U$ is defined in eq.~\eqref{eq:chiral_field} and ellipses denote the higher order terms, and complete Lagrangian to one loop can be found in ref.~\cite{Jiang:2019hgs}. The Lagrangian in eq.~\eqref{eq:HL1} is consistent with the $SU(3)_L\times SU(3)_R$, Lorentz transformations, and the heavy quark symmetry $SU(2)_v$.\par
   The leading order Lagrangian in eq.~\eqref{eq:HL1} can be used to predict the $P^*\rightarrow P\pi$ transitions. Such transitions for $B$ meson are kinematically forbidden however, for the $D$ system, it has the form:
   \begin{align}
       \Gamma\left(D^{*+}\rightarrow D^0 \pi^+\right)&=\frac{g^2}{6\pi F_\pi^2}\vert\vec{p}_\pi\vert^3\,,\\
       \Gamma\left(D^{*+}\rightarrow D^+ \pi^-\right)&=\Gamma\left(D^{*0}\rightarrow D^0 \pi^0\right)=\frac{g^2}{12\pi F_\pi^2}\vert\vec{p}_\pi\vert^3\,.    
   \end{align}
   Using experimental input for these decays, the axial coupling $g$ can be fixed. There are many charmed states which have gained attention over the years as they can not be described by the traditional methods, which require their own review.
  For details on the applications and status of heavy light systems, we refer to refs.~\cite{Casalbuoni:1996pg,Chen:2016spr,Jiang:2019hgs,Meng:2022ozq,Mai:2022eur,Chen:2022asf} and references therein.
   
        \subsection{Soft Collinear Effective Theory, SCET}
   The (light) decay products of heavy (B) mesons typically have a large momentum of order $m_b$, in comparison to $\Lambda_{QCD}$, for instance, in the decay $B \rightarrow K \pi$. The quarks in those fast-moving light mesons are typically on the light cone, collinear with the meson that contains them. Deviations from collinearity are caused by QCD interactions and small, of the order $\Lambda_{QCD}/m_b$ in case of the decay of a $B$ meson. In this situation, the effective theory is constructed `around' those collinear quarks. In an early attempt, Dugan and Grinstein~\cite{Dugan:1990de} constructed a `large energy effective theory (LEET)', to describe the interaction of the high-energy quarks ($E$ around $m_b$) with the soft gluons (energy about $\Lambda_{QCD}$ in an expansion in $q/E$. Since the hadrons also contain collinear gluons, a complete theory must include them too. In refs.~\cite{Bauer:2000ew, Bauer:2000yr},
   Bauer, Fleming, Luke, Pirjol, and Stewart presented a soft collinear effective theory (SCET). A comprehensive description of SCET is found in the book~\cite{Becher:2014oda}, for more recent results and developments, see for instance, refs.~\cite{Bell:2022ott, Goerke:2017ioi}.
   We note that SCET, while originally applied to heavy meson decays, perfectly fits the needs of high energy (jet) physics that is a main part of LHC-physics, see for instance in ref.~\cite{Hoang:2019fze}.
   
   To account for the dominance of the collinear particles, light cone coordinates $p= (p^+, p_\perp, p^-)$ are used. The coordinate basis for motion in the $z$ direction is chosen to
   be $n^\mu = (1,0,0,1)$, $\overline{n}^\mu = (1,0,0,1)$, with $\overline{n}\cdot n =2$ (the coordinates are (t,x,y,z)). The small
   parameter which characterizes the perpendicular components is  $\lambda = p_\perp/{\overline{n}\cdot p}$. The momenta are decomposed according to 
   \begin{align}
       p^{\mu}=\overbrace{\overline{n}\cdot p\frac{n^{\mu}}{2}+\left(p_\perp\right)^\mu}^{\tilde{p}}+n\cdot p\frac{\overline{n}^\mu}{2}=\mathcal{O}(\lambda^0)+\mathcal{O}(\lambda^1)+\mathcal{O}(\lambda^2)\,.
   \end{align}
   This decomposition into large and small components to construct an effective field the theory looks similar to the method of regions, where the different momentum regions are first separated and then treated differently. However, the effective field theory approach allows for systematically including the running of operators or power corrections. The construction of the effective theory then is similar to the theories discussed. SCET also involves three scales like NRQCD. The quantity $\tilde{p}$ now acts as the label to the fields, and the large momenta $\tilde{p}$ are removed by defining:
   \begin{align}
       \psi(x)=\sum_{\tilde{p}}\psi_{n,\tilde{p}}
   \end{align}
   and the derivative $\partial^\mu$ on fields $\psi_{n,p}$ gives dynamical contributions of $\mathcal{O}{(\lambda^2)}$ like in NRQCD. Particle moving along $n^\mu$ have two large components and small components denoted by $\xi_{n,p}$ and $\xi_{\overline{n},p}$ respectively. These are related to $\psi_{n,p}$ by the following relations:
   \begin{align}
       \xi_{n,p}=\frac{\slashed{n}\slashed{\overline{n}}}{4}\psi_{n,p},\quad  \xi_{\overline{n},p}=\frac{\slashed{\overline{n}}\slashed{n}}{4}\psi_{n,p}
   \end{align} satisfying the relations:
   \begin{align}
       \frac{\slashed{n}\slashed{\overline{n}}}{4}\xi_{n,p}&=\xi_{n,p},\quad \slashed{n}\xi_{n,p}=0\,,\\
       \frac{\slashed{\overline{n}}\slashed{n}}{4}\xi_{\overline{n},p}&=\xi_{\overline{n},p},\quad \slashed{\overline{n}}\xi_{\overline{n},p}=0\,.
   \end{align}
  The Lagrangian constructed with the above discussion has the form:
   \begin{align}
       \mathcal{L}_{\text{SCET}}=&\sum_{\overline{p},\overline{p}'}\bigg\lbrace \overline{\xi}_{n,p'}\frac{\slashed{\overline{n}}}{2}\left(i n\cdot D\right)\xi_{n,p}+\overline{\xi}_{\overline{n},p'}\frac{\slashed{n}}{2}\left(\overline{n}\cdot p+i \overline{n}\cdot D\right)\xi_{\overline{n},p}\nonumber\\&\hspace{1cm}+\overline{\xi}_{n,p'}\left(\slashed{p}_\perp+i\slashed{D}_\perp\right)\xi_{\overline{n},p}+\overline{\xi}_{\overline{n},p'}\left(\slashed{p}_\perp+i\slashed{D}_\perp\right)\xi_{n,p}\bigg\rbrace\,,
   \end{align}
where $D_\mu=\partial_\mu-i g T^a A^a_\mu$ is covariant derivative. More details can be found in refs.~\cite{Bauer:2000yr,Becher:2014oda}.
SCET is applied to a large variety of processes with collinear
high-energy particles, not only in decays of heavy mesons but increasingly in very high-energy processes such as at the LHC. For the newest developments, see the latest
SCET conference~\cite{scet2022}.

    \section{Effective theories beyond the standard model}
    \label{SMEFT}
    
    \subsection{The Standard Model Effective Theory}
    
    So far, the standard model has proven to be essentially faultless; apart from a few cosmological phenomena (dark matter, matter-antimatter ratio,...) and alleged
    anomalies in $B$ meson decay~\cite{Alda:2021ruz}, it reproduces all experimental results very precisely. However, it is widely believed that there are more fundamental interactions with a typical energy scale $\Lambda$ which seems considerably higher than $M_W$, as indicated by the absence of discoveries of very heavy particles beyond the top quark, the $W$ and the $Z$ bosons and the Higgs particle at $LHC$. This is reminiscent of the early days of the weak interactions when the 4-Fermi theory  $H_W \sim G_F (\overline{q_L}\gamma^\mu q_L) (\overline{q_L}\gamma_\mu q_L$) was put forward and the $W$-boson entered indirectly only through the Fermi constant $G_F \sim 1/M_W^2$ and its symmetry properties.
    
    Similarly, in order to parameterize physics beyond the standard model, originating from physics at a scale $\Lambda$, one considers effective operators made up of the standard model particles (including the Higgs
    boson and the $W,Z$ bosons with a coupling proportional to powers of $1/\Lambda$,
    \begin{align}
       \mathcal{L} =&\sum_n \frac{1}{\Lambda^n} \mathcal{O}^n\,,
   \end{align}
   where the operators $\mathcal{O}^n$ have dimension $4+n$ (each $\mathcal{O}^n$ consists of many distinct operators, each with an unknown coupling) and are composed out of standard model fields such that the
   total operator is invariant under $SU(3)\times SU(2)\times U(1)$.
    At the lowest order, $(1/\Lambda)^0$, we have just the standard model. At next order, $(1/\Lambda)^1$ there is one operator~\cite{Weinberg:1979sa, Wilczek:1979hc} which violates lepton number. At the next order, $(1/\Lambda)^2$,
    there are nearly 100 operators, see ref.~\cite{Buchmuller:1985jz, Grzadkowski:2010es}. In principle, the task is to determine the unknown couplings by comparing them to suitable experimental results. Given a large number of such couplings, this is a difficult task. This is a very active field, with several strategies to overcome the difficulties.  See ref.~\cite{Brivio:2017vri} for a comprehensive overview. For the newest developments, see the proceedings of
    the 2019 conference on SMEFT-tools~\cite{smeft2019}. This conference will again be held in 2022~\cite{smeft2022}.    

 \subsection{Quantum Gravity}
 
One of the biggest - if not the biggest - unsolved problems in theoretical physics is how to quantize gravity. A modest but important step can be achieved if general relativity is viewed as a field theory. The metric $g_{\mu\nu}$ is promoted as the field, and the effective field theory has the general coordinate invariance of general relativity (GR). Using the fact that the connection, defined as:
\begin{align}
\Gamma_{\alpha\beta}\text{}^{\lambda}=\frac{g^{\lambda\sigma}}{2}\bigg[\partial_\alpha g_{\beta\sigma}+\partial_\beta g_{\alpha\sigma}-\partial_\sigma g_{\alpha\beta}\bigg]\,,
\end{align}
has one derivative, and the curvature, defined in terms of the Riemann tensor ($R_{\mu\nu\alpha\beta}$), given by:
\begin{align}
    R_{\mu\nu\alpha}\text{}^{\beta}=\partial_\mu \Gamma_{\nu\alpha}\text{}^{\beta}-\partial_\nu\Gamma_{\mu\alpha}\text{}^{\beta}-\Gamma_{\mu\lambda}\text{}^\beta\Gamma_{\nu\alpha}\text{}^{\lambda}-\Gamma_{\nu\lambda}\text{}^{\beta}\Gamma_{\mu\alpha}\text{}^\lambda\,,
\end{align}
has two derivatives. The two derivatives present in the Riemann tensor correspond to the powers of energy when evaluated in terms of the matrix elements. It is important to note that the various contractions of the Riemann tensor are coordinate invariant, which is also the symmetry of the low energy theory. Hence, the Lagrangian can be constructed out of various possible contractions of the Riemann tensor, and energy expansion can be naturally constructed including more and more contractions of the Riemann tensor. In particular, Donoghue~\cite{Donoghue:1995cz,Donoghue:scpedia,Donoghue:2012zc}
has shown how a possible extension of general relativity to a theory with quantum degrees of freedom results naturally in an expansion in the theory of gravity, which includes as the 
\begin{align}
    S=\int d^4 x\sqrt{g}\bigg\lbrace\Lambda+\frac{2}{\kappa^2}R+c_1 R^2+c_2 R_{\mu\nu}R^{\mu\nu}+\cdots+\mathcal{L}_{\text{matter}}\bigg\rbrace\,,
\end{align}
where $\Lambda$ is cosmological constant, $R=g^{\mu\nu}R_{\mu\nu}$ and $R_{\mu\nu}=R_{\mu\nu\alpha}\text{}^{\alpha}$ are known as the Ricci scalar and the Ricci tensor, respectively. So far, this theory has found only limited applications, but it may be a guide to correct quantum gravity. A SCET inspired treatment of quantum gravity can be found in refs.~\cite{Beneke:2012xa,Beneke:2021umj,Beneke:2021aip}. For more details, we refer to refs.~\cite{Donoghue:1995cz,Burgess:2003jk,Donoghue:2012zc,Donoghue:scpedia}.

    \section{Miscellaneous items}
    \label{MISC}
    
    In this section, we cover a range of mostly technical topics
    which both feed into effective theories and in whose developments effective theories have played a role.
 \subsection*{Feynman Integral Methods for Effective Field Theories}
   \label{FI}
       When calculating Feynman diagrams, one integrates overall kinematically allowed values of the internal momenta of quarks and gluons. There has been considerable effort in evaluating them to very high orders for precision physics. Many computational as well as theoretical tools have been developed over the years. Many of these developments can be found in the recent book of Weinzierl~\cite{Weinzierl:2022eaz}. In a theory like QCD where the interaction of a gluon with quarks or gluon at $1\GeV$ is very different for an energy scale of several $\GeV$s. The diagrammatic evaluation of any process gets more complicated in these multiple-parameter theories when one goes to higher orders due to the presence of the various scales(masses and momenta) in the loops. It is, therefore, reasonable to divide the integrand into regions and use different rules for the various region. The method of regions~\cite{Beneke:1997zp} is one of the very useful strategies for evaluating Feynman integrals in specific kinematic limits of the mass and momenta. In this technique, the integrand of Feynman diagrams is expanded by identifying the scaling behavior of the ratios of masses and momenta. Although it is not rigorously proven to be correct, it appears to work in all known instances~\cite{Semenova:2018cwy}. Interestingly, the expansion of Feynman diagrams in various regions corresponds to an EFT in the asymptotic limits of the parameters. In some cases, these regions may overlap and need to be systematically subtracted (zero-bin subtraction) following the procedure of Jantzen~\cite{Jantzen:2011nz}. Application of this method in the ChPT was first made by Kaiser and Kaiser and Schweizer~\cite{Kaiser:2006uv}. Now, there exist well-dedicated codes asy.m~\cite{Pak:2010pt}, asy2.m~\cite{Jantzen:2012mw} and ASPIRE algorithm~\cite{Ananthanarayan:2018tog} that can be used to study multi-scale Feynman integrals. For more details, we refer to~\cite{Jantzen:2011nz,Semenova:2018cwy} and references therein. \par
       The Mellin-Barnes (MB) technique is also one of the most commonly used techniques in the literature for the analytic evaluation of the Feynman integrals and has been recently used in the context of ChPT in refs.~\cite{Ananthanarayan:2020xpd,Ananthanarayan:2020fhl,Ananthanarayan:2020ncn}. The two-loop sunset diagrams play a key role in the analytic representation of the masses and decay constants of the pion, kaon, and $\eta$-mesons. These diagrams are calculated using the MB technique in ref.~\cite{Berends:1993ee,Ananthanarayan:2016pos,Ananthanarayan:2017yhz,Ananthanarayan:2018irl,Ananthanarayan:2017qmx} and further used in evaluating some three-loop Feynman diagrams relevant for the QED corrections to $g-2$ of charged leptons in ref.~\cite{Ananthanarayan:2020acj}. The MB technique yields the final expression in terms of generalized hypergeometric functions ($pFq$) and Kamp\'{e} de F\'{e}riet (KdF) series.
       Recently, a geometric method using conic hulls is developed in ref.~\cite{Ananthanarayan:2020fhl} and implemented in the Mathematica package \texttt{MBConicHulls.wl} which allows systematic computation of certain $N$-fold MB integral, and in the case of convergent series case, one can also find the master series which is useful for numerical studies. This technique is used to solve certain non-trivial conformal Feynman integrals in refs.~\cite{Ananthanarayan:2020xpd,Ananthanarayan:2020ncn}.\par
       These ChPT-inspired studies have immensely contributed to finding the new analytic continuations of the Appell Function $F_4$ in terms of the $\text{}_2F_1$ in ref.~\cite{Ananthanarayan:2020xut}. These multivariate hypergeometric functions and their properties, domain of convergences, and linear transformations are studied in mathematics literature~\cite{Bateman:1953,Slater:1966,Exton:1976,Srivastava:1985}. One of the strategies to find the analytic continuation of a multivariate hypergeometric function is to use the known analytic continuations of hypergeometric functions with a lower number of variables. The linear transformation formulae of the one variable Gauss $_2F_1$ function are used to find the analytic continuations of the double variable Appell $F_1$ in~\cite{Olsson64}. This process of finding analytic continuations of hypergeometric series of more than one variable is automated in the Mathematica package \texttt{Olsson.wl}~\cite{Ananthanarayan:2021yar}. The package can also find the domain of convergence of only the double-variable hypergeometric functions. The analytic continuations of the Appell $F_2$ functions are found using the same technique and are used to construct the numerical package \texttt{AppellF2.wl}~\cite{Ananthanarayan:2021bqz}. It can find the numerical value of the Appell $F_2$ function for real values of its arguments (i.e. x, y) and general complex values of the Pochhammer parameters. Some new analytic continuations of Appell $F_4$ are obtained using the known quadratic transformation of the  Gauss $_2F_1$ function~\cite{Ananthanarayan:2020xut}. The linear transformations of the three variable Srivastava $H_C$ function are also found~\cite{Friot:2022dme}. 

\subsection*{Chiral Lagrangians and Ricci Flows}
    
    Right from the early days, the non-linear sigma model provided the fundamental building block for the realization of chiral symmetry.  Whereas for the simplest purposes, these were based on $SU(2)\times SU(2)$ or alternatively on $SO(4)$ general
    theorems for the realization of these symmetries and the Goldstone phenomenon were established for a general group $G$ breaking down to $H$ by Coleman, Wess and Zumino~\cite{Coleman:1969sm}, and Coleman, Callan, Wess, and Zumino~\cite{Callan:1969sn}.  
    Friedan~\cite{Friedan:1980jf,Friedan:1980jm} studied the non-linear sigma model in $2+\epsilon$ dimension where fields $\varphi$ are defined on a manifold $M$ and the coupling are is determined by a Riemannian metric on M. The action has the form:
    \begin{equation}
        S(\varphi)=\Lambda^{\epsilon} \int d x \frac{1}{2} T^{-1} g_{i j}(\varphi(x)) \partial_{\mu} \varphi^{i}(x) \partial_{\mu} \varphi^{j}(x)
    \end{equation}
    where $\lambda$ is short distance cutoff, $T^{-1} g_{i j}(\varphi(x))$ is dimensionless coupling is Riemannian metric on $M$. The renormalization group running of this metric at two-loop was found to be:
    \begin{equation}
        \Lambda^{-1} \frac{\partial}{\partial \Lambda^{-1}} g_{i j}=\beta_{i j}\left(T^{-1} g\right)=-\epsilon\hspace{1mm} T^{-1} g_{i j}+R_{i j}+\frac{1}{2} T\left(R_{i k l n} R_{j k l n}\right)+O\left(T^{2}\right)\,.
        \label{eq:runmat}
    \end{equation}
    and the one-loop $\beta$-function was already calculated by Ecker and Honerkamp~\cite{Ecker:1971xko}. The running of coupling in eq.~\eqref{eq:runmat} is known as Ricci flow introduced by Hamilton~\cite{Hamilton}. The ideas developed by Hamilton were an attempt to solve the long-standing problem of Poincar\'e conjecture.  Perelman published the proof of this conjecture in the three articles~\cite{Perelman1,Perelman2,Perelman3} in 2002-3 where Ricci flow played a key role.
    A detailed explanation of Perelman's proof was published by Morgan and Tian~\cite{MorganT} and Huai-Dong Cao, Xi-Ping Zhu~\cite{Cao}.
    \subsection*{Lattice QCD}
    The non-renormalizable nature of the ChPT results in the increasing numbers of LECs as one goes to higher orders and has to be fixed by inputs from other sources. Most of the LECs can be determined from the experiments or estimated using a large $N_c$ limit of QCD or low energy description of strong interactions. Lattice QCD is one of the candidates at very low energy and has provided numerous inputs and cross-checks over the years. Lattice calculations are performed on finite lattice spacing, finite volume, and unphysical quark masses, and ChPT provides a way to crosscheck, analyze and quantify these effects in the continuum limit. For the brief introduction of the interplay of lattice QCD and the ChPT, we refer to Shanahan~\cite{Shanahan:2016pla} and references therein.\par Among these topics, proton charge radius and the muon $g-2$ anomaly has been of constant interest over the year for their potential to provide hints to new physics beyond the standard model at low energies. Issue of the small charge radius of proton came into the picture in 2010 when the existing value of charge radius $r_p= 0.8775(51)$ fm from CODATA~\cite{Mohr:2012tt} world average using the spectroscopic method and electron-proton scattering was found to be larger than the one obtained from muonic hydrogen $r_p= 0.84184(67)$ fm by Pohl et.al.~\cite{Pohl:2010zza}. Pohl's result later confirmed by CREMA collaboration~\cite{Antognini:2013txn} with $r_p=0.84087(39)$fm. There are various theoretical models for new physics were also studied, and some future experiments are also proposed to get more precise results, but the issue is now believed to be settled and we refer to a very recent review by Gao and Vanderhaeghen~\cite{Gao:2021sml} and Hammer, Mei{\ss}ner~\cite{Hammer:2019uab}, Bernauer~\cite{Bernauer:2020ont}, Peset et. al.~\cite{Peset:2021iul} and references therein for further details. Lattice determinations of form factors are also extensively performed and the results are compatible with existing literature. For details of lattice determination of proton charge radius, we refer to Ishikawa et.al.~\cite{Ishikawa:2021eut} and references therein for details. Lattice methods themselves require their own review to explain various methods developed over the years to extract the parameters of strong interaction. For details, we refer to Golterman~\cite{Golterman:2009kw}, FLAG reviews~\cite{Aoki:2016frl,FlavourLatticeAveragingGroup:2019iem,Aoki:2021kgd}.

\section*{Outlook} \label{sectionOL}

Chiral perturbation theory, ChPT, has proven very fruitful over the last 50 years. It has provided ample
predictions for understanding a great number of experimental results involving the pseudoscalar mesons.
It is still being refined to adapt to new theoretical and experimental results, and there are still many results waiting to be improved. 
ChPT has helped to understand field theory more generally; in particular, it has shed some light on the limited role of renormalizable theories. This direction of research is far from being at its end, and for instance, work devoted to non-renormalizable theories will very likely yield many interesting results~\cite{Ananthanarayan:2018kly} 
Thirdly ChPT has also become a valuable tool to be used in circumstances not thought to be in its
realm. For instance, the calculation of the anomalous magnetic moment of the muon - one of the
crucial calculations in particle physics - has benefited from results obtained by ChPT. We, therefore, believe that chiral perturbation theory, albeit an established and mature technology,
has considerable potential to be improved and gateway to many other developments in particle
physics. 
    \section*{Acknowledgment}
    We thank Dilip K. Ghosh and Sourov Roy for inviting us to write this review.  AK is supported by a fellowship from the Ministry of Human Resources Development, Government of India. We thank Souvik Bera for clarifying remarks and Sumit Banik for help with the manuscript. We also thank the referee for the valuable comments that have improved this article. 

\end{document}